\documentclass[reprint,
superscriptaddress,
amsmath,amssymb,
aps,
]{revtex4-2}

\usepackage{tablefootnote}
\usepackage{booktabs}
\usepackage{tabularx}
\usepackage{tabularray}
\UseTblrLibrary{booktabs}
\usepackage[utf8x]{inputenc}

\usepackage{dcolumn}
\usepackage{amssymb, esint}
\usepackage{physics}
\usepackage{braket}
\usepackage{bm, bbm}
\usepackage{graphicx}
\usepackage[colorlinks,allcolors=blue]{hyperref}
\usepackage{float}
\usepackage{xcolor}
\usepackage{url}
\usepackage[normalem]{ulem}
 \usepackage{mathtools}

\usepackage[nolist]{acronym}

\begin{document}

\title{Analytic Model Reveals Local Molecular Polarizability Changes Induced by Collective Strong Coupling in Optical Cavities}

\author{Jacob Horak}
%\email{jacob.horak@mpsd.mpg.de}
\affiliation{Max Planck Institute for the Structure and Dynamics of Matter and Center for Free-Electron Laser Science, Luruper Chaussee 149, Hamburg 22761, Germany}
\affiliation{The Hamburg Center for Ultrafast Imaging, Luruper Chaussee 149, 22761 Hamburg, Germany}

\author{Dominik Sidler}
\email{dominik.sidler@psi.ch}
\affiliation{Paul Scherrer Institut, 5232 Villigen PSI, Switzerland}
\affiliation{Max Planck Institute for the Structure and Dynamics of Matter and Center for Free-Electron Laser Science, Luruper Chaussee 149, Hamburg 22761, Germany}
\affiliation{The Hamburg Center for Ultrafast Imaging, Luruper Chaussee 149, 22761 Hamburg, Germany}

\author{Thomas Schnappinger}
%\email{thomas.schnappinger@fysik.su.se}
\affiliation{Department of Physics, Stockholm University, AlbaNova University Center, SE-106 91 Stockholm, Sweden}

\author{Wei-Ming Huang}
\affiliation{Department of Physics and Center for Quantum Information Science, National Cheng Kung University, Tainan 70101, Taiwan}
\affiliation{Max Planck Institute for the Structure and Dynamics of Matter and Center for Free-Electron Laser Science, Luruper Chaussee 149, Hamburg 22761, Germany}

\author{Michael Ruggenthaler}
%\email{michael.ruggenthaler@mpsd.mpg.de}
\affiliation{Max Planck Institute for the Structure and Dynamics of Matter and Center for Free-Electron Laser Science, Luruper Chaussee 149, Hamburg 22761, Germany}
\affiliation{The Hamburg Center for Ultrafast Imaging, Luruper Chaussee 149, 22761 Hamburg, Germany}

\author{Angel Rubio}
\email{angel.rubio@mpsd.mpg.de}
\affiliation{Max Planck Institute for the Structure and Dynamics of Matter and Center for Free-Electron Laser Science, Luruper Chaussee 149, Hamburg 22761, Germany}
\affiliation{The Hamburg Center for Ultrafast Imaging, Luruper Chaussee 149, 22761 Hamburg, Germany}
\affiliation{Center for Computational Quantum Physics (CCQ), The Flatiron Institute, 162 Fifth avenue, New York, NewYork 10010, United States of America}

%%%%%%%%%%%%
% Abstract
%%%%%%%%%%%%
\begin{abstract}
Despite recent numerical evidence, one of the fundamental theoretical mysteries of polaritonic chemistry is how and if collective strong coupling can induce local changes of the electronic structure to modify chemical properties. Here we present non-perturbative analytic results for a model system consisting of an ensemble of $N$ harmonic molecules under vibrational strong coupling (VSC) that alters our present understanding of this fundamental question. %The single bare molecular model is composed of one effective electron, which couples harmonically to multiple nuclei. A priori no harmonic approximation is imposed for the inter-nuclear interactions. 
By applying the cavity Born-Oppenheimer partitioning on the Pauli-Fierz Hamiltonian in dipole approximation, the dressed many-molecule problem can be solved self-consistently and analytically in the dilute limit. We discover that the electronic molecular polarizabilities are modified even in the case of vanishingly small single-molecule couplings. Consequently, this non-perturbative local polarization mechanism persists even in the large-$N$ limit. In contrast, a perturbative calculation of the polarizabilities leads to a qualitatively erroneous scaling behavior with vanishing effects in the large-$N$ limit. Nevertheless, the exact (self-consistent) polarizabilities can be determined from single-molecule strong coupling simulations instead. Our fundamental theoretical observations demonstrate that hitherto existing collective-scaling arguments are insufficient for polaritonic chemistry and they pave the way for refined single- (or few-)molecule strong-coupling ab-initio simulations to chemical systems under collective strong coupling.

\end{abstract}


\maketitle

%%%%%%%%%%%%%%%%%%%%%%%%%%%%%%%
% Section
%%%%%%%%%%%%%%%%%%%%%%%%%%%%%%%

\section{Introduction} 

%\textbf{The use of SI units is required}.
Polaritonic chemistry is an emerging field of research at the interface between quantum optics, quantum chemistry and materials science~\cite{ebbesen_2016a,garcia-vidal.ciuti.ea_2021,ebbesen_introduction_2023,ruggenthaler_quantum-electrodynamical_2018,sidler2022perspective,fregoni_theoretical_2022, ruggenthaler_understanding_2023, bhuyan_rise_2023, hirai_molecular_2023,simpkins_control_2023,mandal.taylor.ea_2023,xiang_molecular_2024,sidler_connection_2024}.
By placing matter in a photonic environment, e.g., a Fabry-P\'erot cavity, it has been shown experimentally that chemical and material properties can be modified.
Among others this includes, energy transport~\cite{zhong2016non}, photo-chemical reactions~\cite{hutchison_modifying_2012,munkhbat2018suppression} and also ground-state chemical reactions~\cite{thomas2016ground,lather_cavity_2019,hirai_modulation_2020,fukushima_inherent_2022,ahn_modification_2023,patrahau_direct_2024}.
One of the stunning features of polaritonic chemistry is that these modifications can happen in the "dark", that is, without external illumination.
Instead, chemistry is controlled by the resonances of a photonic structure. For this purpose, a priori intrinsically different control knobs can be identified, such as changing the geometry (e.g., mode volume) or coupling frequency (electronic vs. ro-vibrational) of the cavity. On the matter side, the density, polarizability or composition of the molecular ensemble will affect the strong hybridisation of light and matter inside the optical cavity. Selecting all these parameters wisely  remains a formidable challenge to achieve a desired chemical effect.~\cite{ebbesen_introduction_2023,fregoni_theoretical_2022,svendsen2023theory,svendsen_ab_2024,mandal.taylor.ea_2023}  For a simple and idealized case of a Fabry-P\'erot cavity, a simplified picture emerges in terms of the wavelength of the cavity resonances
\begin{equation}\label{eq:FabryPerotCavity}
    \Lambda = \frac{2 n_{r} L}{m},
\end{equation}
and the effective mode volume $\mathcal{V}$, which scales roughly as $L^3 \mathcal{F}$, where $\mathcal{F}$ is the finesse of the cavity.~\cite{svendsen2023theory} The physical 
length of the cavity is denoted by $L$, the effective refractive index of the medium by $n_r$ and the mode order by $m$. 

%While significant experimental progress has been reported,[REF] the theoretical description is far from complete.[REF]

One critical ingredient to scale up the coupling between light and matter is by \textit{collectively} coupling to a large number $N$ of molecules inside the cavity.  Collective VSC enhances the measured Rabi-splitting by a factor of $\sqrt{N}$.~\cite{thomas2016ground} Consequently, the accurate (microscopic) theory of collective VSC a priori requires to take a very large amount of molecules into account, which makes polaritonic chemistry theoretically extremely challenging.
This is in stark contrast to traditional theoretical chemistry approaches. For instance, in the case of a dilute gas in free space, one can consider a single or few molecules with high computational accuracy  and then perform a statistical treatment of the total ensemble by assuming that the molecules are largely uncorrelated (e.g. canonical ensemble statistics). %In the case of strong coupling, a priori a treatment of the full ensemble including cavity-induced inter-molecular correlations effects seems necessary.
To reach for a computationally manageable complexity of polaritonic chemistry, two opposing theoretical pathways have mainly been explored so far. Either only a single or few molecules are coupled to the cavity field accurately (with a scaled-up coupling constant)~\cite{ruggenthaler.sidler.ea_2023} or, alternatively, large ensemble sizes can be reached with strong simplifications on the coupling and matter description (e.g., Tavis-Cumming-like coupling schemes~\cite{campos2023swinging, mandal.taylor.ea_2023, ruggenthaler.sidler.ea_2023,Borges2024-pn}).

Despite considerable theoretical advances, one of the pressing, still unresolved mysteries of VSC is how (and if) the collective coupling regime can alter individual molecular properties locally as suggested by modified chemical reactions in experiments. Recent numerical evidence suggests that treating the dressed electronic structure problem of the molecular ensemble self-consistently is a crucial ingredient to find cavity-induced microscopic changes under collective VSC. In more detail, a feedback and local-polarization mechanism akin to a spin-glass could been observed,~\cite{sidler_unraveling_2024, schnappinger2023cavity,schnappinger2023ab,sidler_connection_2024} for which computational scaling arguments suggest its persistence in the large-$N$ limit ($N\rightarrow\infty$).~\cite{sidler_unraveling_2024} This is a major step forward in our theoretical understanding, since most models suggest the absence of any local effect.
However, to reach for a more intuitive physical understanding and exploration of the cavity-induced interplay between local and collective properties, a simple exactly solvable ab-initio model is of utter relevance. In this work, we introduce such a model system, for which the Pauli-Fierz Hamiltonian is analytically solvable and provides access to numerous local and collective observables. Most noteworthy, it reveals modified molecular polarizabilities under collective VSC that persist in the thermodynamic limit. 

This manuscript is structured as follows: First we introduce and solve the Pauli-Fierz electronic-structure problem analytically for a dilute gas of $N$ harmonic model molecules. In a second step, cavity-induced local and collective polarization effects are discussed with consequences on the cavity-mode renormalization. Furthermore, a simple recipe is presented to determine molecular polarizabilities ab-initio for complex electronic structures. Afterwards, the classical equations of motions are derived for the nuclei and displacement field coordinates and compared with common approximations in polaritonic chemistry. Eventually, dynamical properties of a canonical ensemble of harmonic CO$_2$ molecules are investigated numerically to verify the analytic predictions. Finally, the analytic formula for the cavity-mode renormalization is tested with accurate single-molecule strong coupling calculations of a realistic CO$_2$ molecule, confirming the chemical relevance of our analytic predictions.

\section{Self-consistent harmonic model for collective VSC}
% - motivate method
% - explain what is to be done
% - do it
% the system hamiltonian
In the following, we investigate collective VSC in a minimal ab-initio molecular setting, which allows an analytic treatment.
For this purpose we look at an ensemble of $N$ identical, non-interacting effective one-dimensional molecules, each one consisting of a single effective electron with negative charge $-Z_e$ and $N_n$ nuclei of mass $M_n$ and with positive charge $Z_n$.
We use atomic units throughout this work.
The corresponding bare matter Hamiltonian is given by
\begin{equation}
    \hat{H}_{\mathrm{m}} = \sum_{i=1}^N \Bigg[\sum_{n=1}^{N_n} \bigg(\frac{\hat{P}_{in}^2}{2 M_n} + \frac{k_{e}}{2}(\hat{R}_{in} - \hat{r}_i)^2\bigg) +  \frac{\hat{p}_{i}^2}{2} + V_i (\mathbf{\hat{R}}_i)\Bigg].
\end{equation}
Nuclear position/displacement and momentum operators are indicated by capital letters, whereas for the $i$-th electron $\hat{r}_{i}$ and $\hat{p}_{i}$ are used, respectively.
The local nucleus-nucleus interaction $V(\mathbf{\hat{R}}_i)$ with $\mathbf{\hat{R}}_i = (\hat{R}_{i1}, ..., \hat{R}_{i N_n})$ will be later parameterized by the force constant $k_n$ (see the specific example given in Eq.~\eqref{eq:harmonicnuclei}), whereas the coupling of the single electron to the nuclei is parameterized by the force constant $k_e$.
The molecular ensemble will be collectively coupled  to a single effective cavity mode $\beta$,~\cite{svendsen2023theory} e.g., of a Fabry-P\'erot cavity. In the length gauge, the Pauli-Fierz Hamiltonian for this system is then given by~\cite{ruggenthaler_understanding_2023}
\begin{equation}\label{eq:pf_dip_h}
%\hat{H} = \hat{H}_\mathrm{matter} + \hat{H}_\mathrm{cavity} + \hat{H}_\mathrm{interaction}
    \hat{H} = \hat{H}_\mathrm{m} + \frac{1}{2} \left[ \hat{p}_\beta^2 + \omega_\beta^2 \left( \hat{q}_\beta - \frac{\hat{X} + \hat{x}}{\omega_\beta} \right)^2 \right] \quad{.}
\end{equation}
\begin{figure}[ht]
\includegraphics[width=\columnwidth]{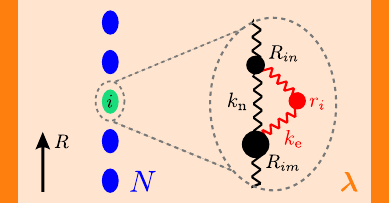}
\caption{Sketch of the setup shows cavity (single mode $\beta$ with frequency $\omega_\beta$ and  polarized parallel to the matter coordinate) with $N$ identical molecules far apart. Each molecule consists of $N_\mathrm{n}$ (different) nuclei and one effective electron of charge $Z_\mathrm{e}$ bound by harmonic interactions. The interactions are associated with force constants $k_\mathrm{n}$ between the nuclei and $k_\mathrm{e}$ between nuclei and effective electron, respectively. The nuclei and electron interact with the cavity via the coupling constant $\lambda$.}
\label{fig:harmonic-setup}
\end{figure}
%Notice, the reason why we consider this a "minimal" molecular model under VSC will become clear later (see Sec. ~\ref{sec:CO2}), where we show that only IR active vibrational modes couple to the cavity and thus the restriction to a single nucleus (atom) would show a qualitatively different behavior. 
We define the polarization operators of the ensemble as
$\hat{X}:=\lambda\sum_{i=1}^N \sum_{n=1}^{N_n} Z_n \hat{R}_{in}$ and $\hat{x}:=-\lambda Z_e \sum_{i=1}^N \hat{r}_{i}$, 
for the nuclei and electrons, respectively. The total dipole of the ensemble couples via $\lambda$ to the effective photon mode with frequency $\omega_\beta$. The corresponding photonic displacement field operator is given by $\hat{q}_\beta$ with canonical momentum $\hat{p}_\beta$.
We note that for ground-state chemical reactions the effective single-mode approximation is expected to capture all the essential physics~\cite{schafer2023machine,svendsen2023theory}.
The main effect of a proper continuum of modes amounts to an extra dissipation channel as this leads to the radiative decay of the coupled system~\cite{flick.welakuh.ea_2019,ruggenthaler.sidler.ea_2023}.

The vectorial photon-matter coupling $\bm{\lambda}_\beta = \bm{\varepsilon}_\beta \lambda$ depends on the mode polarization vector $\bm{\varepsilon}_\beta$ and the coupling constant~\cite{ruggenthaler.sidler.ea_2023} 
\begin{equation}\label{eq:lambda-def}
    \lambda = \sqrt{\frac{4 \pi}{\mathcal{V}}},
\end{equation}
where $\mathcal{V}$ corresponds to the effective mode volume.
Here we have assumed for simplicity that all effective one-dimensional molecules are perfectly aligned with respect to the polarization vector $\bm{\varepsilon}_\beta$.
However, qualitatively similar results are expected for randomly oriented molecules as has previously been shown numerically in Ref.~\cite{sidler_unraveling_2024}.

If we would restrict to purely harmonic nucleus-nucleus interactions (as we do later in Sec.~\ref{sec:CO2}) we could in principle consider the full quantum dynamics by merely solving the corresponding classical equations of motion due to the harmonic nature of the model.~\cite{schnappinger2023ab}
However, we here do not assume this and we connect to the cavity Born-Oppenheimer approximation~\cite{flick_cavity_2017,ruggenthaler.sidler.ea_2023}, by assuming that the electrons adapt instantaneously on the time-scale of the nuclei and displacement coordinate.
Thus we partition our polaritonic problem into two coupled sub-problems, one for the electrons and one for the nuclear and displacement degrees of freedom.
The Hamiltonian operator of the coupled nuclear-photon degrees of freedom on the $l$-th electronic potential-energy surface is given by
\begin{align}
    \hat{H}&^{\mathrm{npt},l} := \sum_{i=1}^N \Bigg[ \sum_{n=1}^{N_n}\bigg(\frac{\hat{P}_{in}^2}{2 M_n}+\frac{k_{e}}{2}\hat{R}_{in}^2\bigg) +V_i (\mathbf{\hat{R}}_i)\Bigg] \notag \\
    &+\frac{\hat{p}_\beta^2}{2}
    +\frac{\omega_\beta^2}{2}\Big(\hat{q}_\beta-\frac{\hat{X}}{\omega_\beta} \Big)^2 + \underbrace{\bra{\Psi_l}\hat{H}^\mathrm{e}(\mathbf{R}, q_\beta) \ket{\Psi_l}}_{= E^{\mathrm{e}}_{l}(\mathbf{R},q_\beta)}. \label{eq:CBO_npt}
\end{align}
The corresponding $N$-electron Hamiltonian operator is given by
\begin{align}
 \hat{H}^{\mathrm{e}}(\mathbf{R}, q_\beta) := \sum_{i=1}^N &\Bigg[\frac{\hat{p}_{i}^2}{2} - \sum_{n=1}^{N_n}k_{e} R_{in}\hat{r}_i+N_n k_{e}\frac{\hat{r}_i^2}{2}\Bigg] \notag \\
 &+\bigg(\frac{1}{2} \hat{x}^2+\hat{x} X-\omega_\beta  \hat{x} q_\beta\bigg), \label{eq:Helectron}
\end{align}
where the electrons only interact due to the presence of the (effective) cavity mode. 
The electronic Hamiltonian depends only parametrically on all the nuclei positions and displacement field coordinates, written compactly as $(\mathbf{R},q_\beta)$.

Assuming the dilute gas limit (i.e., no electron-electron interaction or setting the integrals $\braket{\psi_i | \hat{O} | \psi_k}_{i \neq j} = 0$ for any operator $\hat{O}$), the Hartree-Fock wave function reduces to a Hartree product $\Psi = \psi_1 \otimes \psi_2 \otimes \dots \otimes \psi_i \otimes \dots \otimes \psi_{N}$ of $N$ single-molecule electronic wave functions $\psi_i$.
These are determined by the following coupled $N$ Hartree equations~\cite{sidler_unraveling_2024, schnappinger2023cavity}
\begin{align}
 &\bigg(\frac{\hat{p}_{i}^2}{2}-\sum_{n=1}^{N_n} k_e R_{in}\hat{r}_i+N_n k_e \frac{\hat{r}_i^2}{2} \label{eq:CBO_electronic} \\
 &+ \Big(X - q_\beta \omega_{\beta} + \sum_{j \neq i}^{N} \braket{\psi_j | \hat{x}_{j} | \psi_j} \Big) \hat{x}_{i} +
\frac{\hat{x}_{i}^2 }{2}  \bigg) \psi_i = 	\varepsilon_i \psi_i \quad{.}\nonumber
\end{align}
This model resembles the numerical approach in Ref. \cite{castagnola_changes_2024} to mimic collective effects in 
Note that in analogy to Hartree-Fock theory,~\cite{szaboModernQuantumChemistry1996a} the total electronic energy is not just the sum of the single molecule energies but one needs to subtract the energy of the two-electron operator to cancel double counting
\begin{eqnarray}
  E^{\mathrm{e}} = \sum_i^N \epsilon_i - \frac{1}{2} \Big( \braket{x}^2 - Z_e^2 \lambda^2 \sum_i^N \braket{r_i}^2 \Big).  
\end{eqnarray}
Thanks to our harmonic matter description, the $N$-electron problem given in Eq.~\eqref{eq:CBO_electronic} can be solved analytically. By defining 
\begin{eqnarray}
    \mu_i:= Z_e \lambda \sum_{j \neq i}^{N} \braket{\psi_j | \hat{r}_j | \psi_j},
\end{eqnarray}
the $i$-th electron problem corresponds to a shifted harmonic oscillator parameterized by $(\mathbf{R},q_\beta)$ as
\begin{align}
    \bigg[&\frac{\hat{p}_i^2}{2} + \nu_{1,i} \hat{r}_i + \frac{\nu_2^2}{2}\hat{r}_i^2\bigg]\psi_i(r_i) = \epsilon_i \psi_i(r_i),\label{eq:local_shift_osc}\\
    \nu_{1,i} &:= -k_{e}\sum_{n=1}^{N_n}R_{in} +Z_e \lambda (- X + \omega_\beta q_\beta  + \mu_i) \\
    \nu_2 &:= \sqrt{\lambda^2 Z_e^2 + N_n k_{e}} \label{eq:shifted-HO}
\end{align}
The shifted harmonic oscillator energies are then
\begin{align}
    \epsilon^l_i &= \Braket{\hat{b}_i^\dagger \hat{b}_i +\frac{1}{2} -\eta_i^2(\mathbf{R},q_\beta)}_l \nu_2 \notag \\
    &= \left( l + \frac{1}{2} - \frac{\nu_{1,i}^2}{2 \nu_2^3} \right) \nu_2 \label{eq:shifted-HO-energies}%\\
    %&\eta_i (\boldsymbol{R},q_\beta) =\frac{\nu_{1,i}}{\sqrt{2 \nu_2^3}} \\
    %&\hat{b}_i^\dagger = \hat{a}_i^\dagger + \eta_i \, , \quad
    %\hat{b}_i = \hat{a}_i +\eta_i 
\end{align}
where $\eta_i (\mathbf{R},q_\beta) =\nu_{1,i}/\sqrt{2 \nu_2^3}$ and $\hat{b}_i^\dagger = \hat{a}_i^\dagger + \eta_i$, \
    $\hat{b}_i = \hat{a}_i +\eta_i$.   The ladder operators of the original quantum harmonic oscillator are given by $\hat{a}_i^\dagger = \sqrt{\nu_2 / 2}(\hat{r}_i-i \hat{p}_i/\nu_2)$ and $\hat{a}_i = \sqrt{ \nu_2 / 2}(\hat{r}_i + i \hat{p}_i / \nu_2)$ with $\hat{r}_i=\sqrt{1/2\nu_2}(\hat{a}_i^\dagger + \hat{a}_i)$. In other words, the solutions of the shifted quantum harmonic oscillator are connected to the original harmonic oscillator via the unitary displacement operator $\hat{U}_i = e^{- \eta_i \left(\hat{a}_i^\dagger - \hat{a}_i \right)}$.

Since the instantaneous dipoles of the electrons couple to the cavity field we determine the expectation value of the local and the total electric dipole for later reference. When calculating the electronic position expectation value of the $i$-th electron in the ground state, we find a recursive dependency on the position of all other electrons due to the dipole-dipole interaction term, i.e.,
\begin{align}
    \langle \hat{r}_i\rangle_l &=\langle \hat{r}_i\rangle_0 = -\sqrt{\frac{1}{2\nu_2}}(\eta_i+\eta_i^*) = -\frac{\nu_{1,i}}{\nu_2^2} \label{eq:r_i_base}\\
    %&= \frac{1}{\nu_2^2} \bigg[
    %    k_e \sum_{n=1}^{N_n}R_{in} + Z_e \lambda \bigg(X - \omega_\beta q_\beta -   \mu_i \bigg) 
    %\bigg] \notag \\
    &=\frac{1}{N_n}\bigg(\sum_{n=1}^{N_n}R_{in} +\frac{\lambda Z_e }{ k_e}\Big(X +\langle x\rangle_0-  \omega_\beta q_\beta\Big)\bigg) \label{eq:r_i}
    \\&=\frac{1}{N_n}\bigg(\sum_{n=1}^{N_n}R_{in} -\frac{E_\perp Z_e}{ k_e}\bigg)
    \label{eq:rietrans}
\end{align}
where we have implicitly used $\mu_i = Z_e \lambda\sum_{j}^N \langle\hat{r}_j\rangle_0 - \langle\hat{r}_i\rangle_0$ in the second line.
The last step introduces the definition of the transverse electric field
\begin{eqnarray}
    E_\perp= \underbrace{\lambda \omega_\beta q_\beta}_{4\pi D} - \underbrace{\lambda (X +\langle x\rangle)}_{4\pi P} \label{eq:eperpdef} 
\end{eqnarray}
in the length gauge in terms of the displacement $D$ and polarization fields $P$.~\cite{sidler.ruggenthaler.ea_2022, ruggenthaler.sidler.ea_2023}
We note that if we only couple to the displacement field without including the self-consistent polarization of the system ($E_\perp = 4 \pi D$), then we  find a different shifted oscillator as detailed in Sec.~\ref{sec:approximations}.
%This assumption, obtained by dropping the quadratic dipole self-energy (dse) terms in Eq.~\eqref{eq:pf_dip_h} and only keeping $-\omega_{\beta} \hat{q}_{\beta} (\hat{X} + \hat{x})$, is intrinsic of various approximations such as the Tavis-Cummings model~\cite{sidler.ruggenthaler.ea_2022, ruggenthaler.sidler.ea_2023}. \revJH{Not so sure about the Tavis-Cummings thing. Hopfield is µ model but more extreme? Without dse we will call it linear}
%\tbdel{We would obtain the same result if we replaced the displacement field by an external electric field instead. In more detail, via $-\omega_{\beta} \hat{q}_{\beta} (\hat{X} + \hat{x}) \rightarrow -E_{\rm ext}(\hat{X}+\hat{x})|e|/\lambda$ we could just replace $\omega_{\beta} q_{\beta} \lambda \rightarrow E_{\rm ext} |e|$ in Eq.~\eqref{eq:localdipolenoselfenergy}. \revJH{why is this interesting?}
%We note that this is only possible because we have a harmonic potential, which becomes arbitrarily strong as we move away from the minimum of the binding potential. For more realistic (Coulombic) potentials, which approach a finite value for $|r_i| \rightarrow \infty$, such dipolar couplings would become problematic since no ground state would exist in an ab-initio description~\cite{rokaj_lightmatter_2018,schafer_relevance_2020}.}

Now, we can perform the $\lambda$-weighted summation over all $N$ molecules from Eq. \eqref{eq:r_i_base} to derive analogously the exact relation
\begin{align}\
    \braket{x} = \big(1 - &\gamma^{2} \big) \bigg[
        - \frac{k_e}{\lambda Z_e N} \sum_{i}^N \sum_{n=1}^{N_\mathrm{n}} R_{i n}
        - X 
        + \omega_\beta q_\beta
    \bigg]\label{eq:relationx} \\
    &\gamma^2(N,\lambda) =\frac{1}{1+\lambda^2 N \alpha_i } %\frac{N_\mathrm{n} k_e}{\lambda^2 Z_e^2 N + N_\mathrm{n} k_e}
    , \quad 0< \gamma^2\leq 1.\label{eq:gamma2}
\end{align}
%{\color{green} (Is this the Lorentz model if we write $\alpha_{\rm Lorentz} = N \alpha_i \left(1/(1+N \lambda \alpha_i)\right)$ due to micro-macro connection? So we might see that beyond harmonic we will have all the higher orders.)} \revDS{I would keep that for a future publication, when we can investigate the impact on the condensed phase.} 
Here we have introduced the scaling constant $\gamma^2(N,\lambda)$, which depends on the single molecular (bare matter) polarizability $\alpha_i=Z_e^2/(N_n k_e)$ (see discussion in Sec. \ref{sec:pol}).  
%Note that we have implicitly assumed that we are dealing with a finite number of molecules $N$ as well as that the individual summands are finite. %Relaxing this assumptions will be relevant to discuss  thermodynamic limit properties and will be the focus of a future publication. 
Eventually, using Eq.~\eqref{eq:relationx} one can rewrite the transverse electric field, given in Eq.~\eqref{eq:eperpdef}, solely in terms of nuclear and displacement field coordinates. This will be convenient to  find self-consistent equations of motion for the combined nuclei-displacement subsystem and to have access to non-perturbative  polarizability features.

% -------------------------------- FEATURES ---------------------------
\section{Features of ensemble self-consistency}
We now have a self-consistent ab-initio model to investigate VSC in the collective regime.
Having access to analytic expressions for all electronic expectation values and therefore its derivatives, we can make use of the Hellmann-Feynman theorem to easily investigate different properties $O$ as $\frac{\mathrm{d} \braket{\hat{H}}} {\mathrm{d} O} = \braket{\frac{\mathrm{d}}{\mathrm{d} O} \hat{H}}$.
First, we explore the polarizabilities and re-scaling of the cavity frequency.
Then, we look at the equation of motion for the nuclei under VSC and show the effect of common (non-self-consistent) approximations.
%In the following, we will look at the equation of motion of the displacement coordinate, the nuclei and the polarizability.
%This is expected as our model is only a simplification of self-consistent ab-initio CBOA treatment of VSC.

\subsection{Modified Polarizabilities Under Collective VSC}
\label{sec:pol}
Molecular polarizabilities are important to understand and probe molecular properties with light. Moreover, they also influence how molecules interact. For example, London dispersion forces are a key element to understand solvation processes~\cite{atkins2014PhysicalChemistryQuanta,philbin_molecular_2023} and they have successfully been altered in experiment.~\cite{patrahau_direct_2024} 
The standard definition of the (static) electronic polarizability $\alpha$ of a \textit{single} molecule or an ensemble of uncoupled $N$ molecules considers how the respective dipole moment $\hat{d}$ responds to an external electric field $E_\mathrm{ext}$, described by $\hat{H}_m - \hat{d} E_\mathrm{ext}$ ~\cite{atkins2011molecular} 
\begin{equation}
\alpha = \frac{\partial \braket{d}}{\partial E_\mathrm{ext}} = \sum_i^N Z_e \frac{\partial \braket{\hat{r}_i}}{\partial E_\mathrm{ext}}.\label{eq:polarizability_def}
\end{equation}
The electronic dipole operator is given as $\hat{d} = Z_e \sum_i^N \hat{r}_i$.
In case of uncoupled (bare) harmonic molecules (i.e., only probing $\hat{H}_\mathrm{M}$) the collective polarizability can either be calculated self-consistently or perturbatively (see Sec.~\ref{sec:linpol}), yielding the identical analytical result
\begin{equation}\label{eq:polarizability}
    \alpha = N \underbrace{\frac{Z_{e}^{2}}{{N_\mathrm{n}} k_{e}}}_{\alpha_i},%\label{eq:pol_col_mat}
\end{equation}
where $\alpha_i$ is the single-molecule polarizability. 
%corresponds to the molecular polarizability for the single-molecule case $N=1$.
Notice that $\alpha_i$ is defined by a local perturbation within the uncoupled (bare) matter ensemble, i.e., by $\hat{H}_m - \hat{d}_i E^{i}_\mathrm{ext}$, and the ensemble polarizability is merely the single-molecule polarizability scaled linearly by the number of molecules $N$. This simple linear relation also holds if we ask, for instance, what would be the response of any subset of molecules of our ensemble to a small perturbation.
%, which confirms the linear scaling with $N$, as shown in Eq.~\eqref{eq:polarizability}. 
%More realistic electronic (Coulomb instead of harmonic potentials), bare matter polarizabilities are typically calculated by perturbation theory for a single-molecule, since the Hamiltonian, which includes the external field in dipole approximation, becomes unstable in the basis set limit.~\cite{rokaj_lightmatter_2018,rokaj_lightmatter_2018,schafer_relevance_2020}  %In principle, in presence of a cavity mode, this theoretical inconsistency is lifted by the dipole self-energy term which makes a non-perturbative calculation of $\alpha$ feasible. This self-consistent treatment will turn out to be essential to capture the correct physics, as we will detail subsequently.  
%In more, we will show later in Sec. \ref{sec:approximations} that a perturbative determination of cavity-mediated polarizabilities yields identical results for the harmonic bare matter problem, which will no longer be the case under VSC. 
%We find that this bare electronic polarizability is in agreement with the one obtained from comparison with the red shift via the refractive index from Eq.~\eqref{eq:ref-classical}.
%
%Having direct access to non-perturbative polarizabilities, our harmonic setup is an ideal model to investigate the cavity-induced feedback effects between the molecules upon a small external perturbation. Indeed, as we will see subsequently, the perturbative and self-consistent treatment of an external perturbation leads to a qualitatively different scaling behaviour and thus different physics.

For our self-consistent solution of the ensemble, we can check whether this simple linear relation, which reduces everything back to the single-molecule response, is recovered. To check this we consider a few different polarizabilities, specifically the polarizability of the full ensemble $\tilde{\alpha}$, the polarizability of the $i$-th molecule if the full ensemble is perturbed $\tilde{\alpha}_i$, the polarizability of the full ensemble if only the $j$-th molecule is changed $\tilde{\alpha}^{j}$ and finally the polarizability of the $i$-th molecule if the $j$-th molecule is perturbed locally $\tilde{\alpha}_i^j$. Mathematically these different polarizabilities are obtained by
\begin{eqnarray}
   \tilde{\alpha} = \frac{\partial \braket{d}}{\partial E_\mathrm{ext}} \ \mathrm{with } \ \hat{H}_{\rm tot}=\hat{H} - \hat{d} E_\mathrm{ext}\\
   \tilde{\alpha}_i = \frac{\partial \braket{d_i}}{\partial E_\mathrm{ext}} \ \mathrm{with } \ \hat{H}_{\rm tot}=\hat{H} - \hat{d} E_\mathrm{ext}\\
   \tilde{\alpha}^j = \frac{\partial \braket{d}}{\partial E_\mathrm{ext}} \ \mathrm{with } \ \hat{H}_{\rm tot}=\hat{H} - \hat{d}_j E_\mathrm{ext}^{j}\\
   \tilde{\alpha}_i^j = \frac{\partial \braket{d_i}}{\partial E_\mathrm{ext}} \ \mathrm{with } \ \hat{H}_{\rm tot}=\hat{H} - \hat{d}_j E_\mathrm{ext}^{j}
\end{eqnarray}
In dipole approximation the perturbations preserve the harmonicity of our dressed many-electron problem, and thus $\hat{H}_{\rm tot}$ can be solved analytically and the derivatives are straightforward to calculate. The corresponding, self-consistently calculated polarizabilities in a cavity are then found to be (see App.~\ref{sec:polder} for details) 
\begin{eqnarray}
   \tilde{\alpha}(N,\lambda) &=& \alpha \gamma^2(N,\lambda)\label{eq:coll_polarizability_sc},\\
   \tilde{\alpha}_i(N,\lambda) &=& \alpha_i \gamma^2 (N,\lambda),\\
   \tilde{\alpha}^j(N,\lambda) &=& \alpha_i\gamma^2(1,\lambda)\big[1-(N-1)\lambda^2 Z_e^2\big]\\
   \tilde{\alpha}_i^j(N,\lambda)  &=& \begin{cases*}
                     \alpha_i\gamma^2(1,\lambda) & if  $i=j$  \\
                      -Z_e^2\lambda^2 \alpha_i\gamma^2(1,\lambda) & else.
                 \end{cases*}\label{eq:local_local}
\end{eqnarray}
As we see, the simple linear scaling is not recovered for this set of polarizabilities. To some extend this is not a surprise, since we are considering a coupled ensemble of molecules. Yet this result has profound implications for polaritonic chemistry. These implications become specifically striking if we assume the naive Tavis-Cumming scaling
%If we now assume the usual Tavis-Cummings scaling of the local coupling parameter (which has been confirmed numerically for our self-consistent setup in Ref. \citenum{sidler_unraveling_2024}) 
\begin{equation}\label{eq:collective-coupling-strength}
    \lambda_{\rm TC} = \frac{\lambda_\mathrm{col}}{\sqrt{N}},
\end{equation}
which neglects that physically there is a largest length scale over which we can couple different molecules via a cavity mode.~\cite{svendsen2023theory} In other words, the mode volume $\mathcal{V}$ of Eq.~\eqref{eq:lambda-def} sets a maximal volume into which we can put our ensemble. Consequently, if we choose a specific molecular density $N/V$, where $V$ can be viewed as the (arbitrary) quantization volume of our theoretical description, and choose $V>\mathcal{V}$, we are effectively describing a different cavity situation. Thus the naive $N \rightarrow \infty$ limit should be handled with care. Disregarding this issue for the moment, we can still see what taking such a limit entails. For the few examples of polarizabilties we find
\begin{eqnarray}
    \tilde{\alpha}(N,\lambda_{\rm TC})/N &\rightarrow& \alpha_i \gamma^2(1,\lambda_\mathrm{col}),\label{eq:coll_polarizability_sc_lim_glob}\\
   \tilde{\alpha}_i(N,\lambda_{\rm TC}) &\rightarrow& \alpha_i \gamma^2 (1,\lambda_\mathrm{col}),\label{eq:coll_polarizability_sc_lim_loc}\\
   \tilde{\alpha}^j(N,\lambda_{\rm TC}) &\rightarrow& (1-\lambda_\mathrm{col}^2 Z_e^2)\alpha_i,\label{eq:col_loc_resp}\\
   \tilde{\alpha}_i^j(N,\lambda_{\rm TC}) &\rightarrow& \begin{cases*}
                    \alpha_i & if  $i=j$  \\
                      0 & else
                 \end{cases*}.\label{eq:pol_loc_loc}
\end{eqnarray}
When looking at Eq.~\eqref{eq:pol_loc_loc}, one could be tempted to conclude that locally the molecules in an ensemble are not modified and are all statistically independent. Based on this we would obtain as the polarizability of the full ensemble the linear relation of Eq.~\eqref{eq:polarizability}. This is clearly wrong, since we should obtain Eq.~\eqref{eq:coll_polarizability_sc_lim_glob}, which is the exact expression just normalized by the number of molecules $N$. The same is true also for the polarizability of Eq.~\eqref{eq:col_loc_resp}. So, what went wrong?

The answer is as simple as it is insightful for the case of polaritonic chemistry. An individual molecule in the cavity-coupled ensemble is \textit{not} statistically independent from the rest. Consequently, if we want to know how molecules in the ensemble react, we need to keep track of all the different possibilities the cavity lets them talk to each other. And although the local-local relation of Eq.~\eqref{eq:local_local} is only slightly modified by the cavity, a single molecule in the ensemble is also modified if any number of other molecules is perturbed. Since the number of all possible coupled perturbations increases as we increase the number of molecules, the self-consistent solution stays modified even for $N \rightarrow \infty$. Thus using the limit of Eq.~\eqref{eq:pol_loc_loc} and then making statements about the properties of the ensemble is inconsistent.

We consider the fact that the polarizabilities are modified even in the large-$N$ limit the first quintessential result of this work. Indeed, under VSC polarizabilities  can depend on the collective Rabi splitting. This collectively-induced local effects can only appear when self-consistently solving the dressed electronic-structure problem.
Notice further that the collective dipole response in Eq.~\eqref{eq:col_loc_resp} switches sign for $\lambda_\mathrm{col}^2 Z_e^2\rightarrow 1$, i.e., the collective medium polarizes opposite to the locally applied electric field. 
To summarize, we have seen that collective VSC changes global and local polarizabilities. Looking only at the single-molecule response \textit{without} taking into account the statistical dependence on the other molecules can lead to the wrong conclusion that in the large-$N$ limit the ensemble is not modified. Let us see next, how a perturbative treatment of the coupled ensembles performs, which is a standard ansatz based on the Tavis-Cumming-scaling argument.

 %$\frac{\alpha_i N_\mathrm{n} k_e}{\lambda_\mathrm{col}^2 Z_e^2 + N_\mathrm{n} k_e}$.

%Notice further that for our harmonic effective electronic structure, Eqs. \eqref{eq:polarizability} and \eqref{eq:coll_polarizability_sc} become exact, i.e. is a non-perturbative solution.

\subsubsection{Polarizabilities from Perturbation Theory\label{sec:linpol}}

Treating $\hat{d}_j E_\mathrm{ext}^{j}$ or $\hat{d} E_\mathrm{ext}$ in perturbation theory, the above four cases reduce to effectively two cases. The local polarizability can be calculated from~\cite{atkins2011molecular} 
\begin{eqnarray}
    \tilde{\alpha}^{\rm pert}_i(N,\lambda)= -2\sum_{l\neq 0} \frac{\bra{0}Z_e \hat{r}_i\ket{l}\bra{l}Z_e \hat{r}_i\ket{0}}{\epsilon_i^0-\epsilon_i^l}
\end{eqnarray}
as (see App.~\ref{sec:polder} for details)
\begin{eqnarray}
    \tilde{\alpha}^{\rm pert}_i(N,\lambda)&=& \alpha_i\gamma^2(1,\lambda)\nonumber\\
    &\overset{N\gg 0}{\rightarrow }& \alpha_i, \label{eq:pert_loc}
    \end{eqnarray}
where we have used that the shift of the harmonic oscillator does not affect the overlap matrix elements, i.e., the standard bare matter expressions from Ref.~\citenum{atkins2011molecular} can be used. The only difference to a bare matter ensemble emerges from the cavity-modified renormalization of the electronic excitation energy, i.e., $\nu_2^2=N_n k_e \mapsto \lambda^2 Z_e^2+ N_n k_e$ within a cavity.

For the collective perturbation one simply finds 
\begin{eqnarray}
    \tilde{\alpha}^{\rm pert}(N,\lambda)&=& -2\sum_{l\neq 0} \frac{\bra{0}\sum_i^N Z_e \hat{r}_i\ket{l}\bra{l}Z_e \sum_i^N\hat{r}_i\ket{0}}{\epsilon_i^0-\epsilon_i^l} \nonumber \\
    &=& N  \tilde{\alpha}^{\rm pert}_i(N,\lambda). \label{eq:pert_glob}
\end{eqnarray}
We note that $\ket{0}$ corresponds to the collective electronic ground state formed by the product state of the $\ket{\psi_i}$ wave functions, which are the local ground states with $l=0$. Similarly the collective excited states $\ket{j}$ labels all possible excited state combinations. However, only very few combinations can contribute, as shown in App.~\ref{sec:polder}. 
Eventually, from perturbation theory and by assuming the usual Tavis-Cummings scaling of Eq.~\eqref{eq:collective-coupling-strength} we find that the polarizabilities of the molecular ensemble are the same inside and outside the cavity. This, however, again contradicts the exact results obtained in Eqs.~\eqref{eq:coll_polarizability_sc_lim_glob} - \eqref{eq:pol_loc_loc}. The reason being that perturbation theory is not taking into account all possible ways the cavity let's the molecules of the ensemble talk to each other.

Interestingly, the perturbative result in Eqs.~\eqref{eq:pert_loc} and \eqref{eq:pert_glob} agree with the exact self-consistent solution in Eqs.~\eqref{eq:coll_polarizability_sc_lim_loc} and \eqref{eq:coll_polarizability_sc_lim_glob}, provided that one assumes single-molecule strong coupling with a scaled coupling constant $\lambda\rightarrow\lambda_{\rm col}$. The resulting relations 
\begin{align}
\boxed{
    \tilde{\alpha}_i(N, \lambda)=\tilde{\alpha}_i(1,\lambda_\mathrm{col})= \tilde{\alpha}_i^{\rm pert}(1,\lambda_{\rm col})\neq \tilde{\alpha}_i^{\rm pert}(N,\lambda)}\label{eq:wrong_scaling}
\end{align}
we consider the second quintessential results or this work. Indeed, Eq.~\eqref{eq:wrong_scaling} provides a practical recipe to determine the single molecule-response of a large ensemble under collective VSC. In particular, it provides access to collectively-dressed polarizabilities of complex electronic structures by solving $\tilde{\alpha}_i(1,\lambda_\mathrm{col})=\tilde{\alpha}_i^{\rm pert}(1,\lambda_{\rm col})$ for a single molecule strongly coupled to a cavity. We note that the perturbative treatment here refers to the external field \textit{not} the coupling to the scaled-up cavity mode $\lambda_{\rm col}$.
%This can be done, either by solving a strongly coupled single-molecule subject to an external field, or might be done perturabtively on top of a single-molecule strong coupling ground-state calculation.  The perturbative single-molecular strong coupling approach avoids some theoretical subtleties. %that  the non-perturbative inclusion of an external electric field can cause.  
%In more detail, for realistic electronic structures (Coulombic) a non-perturbative inclusion of the external electric field introduces severe theoretical inconsistencies, which are not present in our harmonic model, however. For Coulombic potentials, the ground state would not even exist for the bare matter problem in the basis set limit[REF]. If coupled to a cavity, a ground state  would only exist thanks to the dipole-self-energy term of the cavity, provided that the polarization axis of the cavity and the external electric field perturbation are exactly aligned. Still the resulting numerical polarizability would remain ad-hoc and would be physically questionable.  Rigorously, these theoretical inconsistencies could only be avoided by including the source of the external electric field (e.g., charged plates) to the entire system and performing a full self-consistent ab-initio calculation. However, this would likely be computationally very demanding even for a single molecule in a cavity and thus the perturbative single-molecular strong coupling approach seems the method of choice. 
Notably, Eq.~\eqref{eq:wrong_scaling} confirms analytically that accurate single or few molecular strong coupling ab-initio methods~\cite{ruggenthaler.flick.ea_2014,tokatly_2013a,ruggenthaler2015ground,ruggenthaler_understanding_2023,schafer_making_2021,pellegrini.flick.ea_2015,sidler_chemistry_2020,haugland_coupled_2020,mordovina_polaritonic_2020,rivera_variational_2019,ahrens_stochastic_2021,schnappinger2023cavity,flick_cavity_2017,flick_atoms_2017,tancogne-dejean.oliveira.ea_2020, schafer_making_2021,li_collective_2021,szidarovszky2021nonadiabatic,fabri2021born}, which were developed over the past decade, are indeed practically relevant to determine cavity-modified changes and thus chemistry under collective strong coupling conditions. In particular, it confirms that simple scaling arguments\cite{martinez-martinez_triplet_2019,du_vibropolaritonic_2023} based on standard quantum-optical models (e.g. Tavis-Cummings) are inadequate\cite{ebbesen_2016a} to judge the experimental relevance of accurate single-molecular strong coupling ab-initio results.
Notice again, all our self-consistent derivations assume the gas phase approximation. Possible consequences on the condensed phase, e.g., cavity-induced modifications to the Clausius-Mossotti relations, will be the focus of a future publication. 

%\begin{eqnarray}
%    \tilde{\alpha}^{\rm pert}&=& N  \tilde{\alpha}^{\rm LR}_i\overset{N\gg 0}{\rightarrow } \alpha
%    \end{eqnarray}

\subsection{Cavity Frequency Renormalization}
Collective strong coupling does not only change molecular polarizabilities, but it also
renormalizes the cavity frequency. This can be directly seen when investigating the classical equations of motion for the nuclei and displacement field coordinates. Using the Hellman-Feynman theorem, we find for the dynamics of the displacement coordinate 
\begin{align}\label{eq:qddot-base}
\ddot{q}_\beta &=  -\Braket{\frac{\mathrm{d} \hat{H}}{\mathrm{d} q_\beta}}_l = - \omega_\beta^2 \bigg(q_\beta - \frac{X + \braket{ \hat{x}}}{\omega_\beta} \bigg) \\
&= -\gamma^2 \omega_\beta^2 q_{\beta} + \gamma^2 \omega_\beta X - \big(1- \gamma^2 \big) \frac{k_e}{N Z_e \lambda} \mathcal{R}. \label{eq:qddot}
\end{align}
The nuclear sums are abbreviated as $\mathcal{R} = \sum_{in}^{N N_\mathrm{n}} R_{in}$.
This reveals a renormalization of the photon mode frequency by
\begin{equation}\label{eq:selfconsistentfrequency}
    \omega_{\beta} \rightarrow \tilde{\omega}_{\beta} =\gamma \omega_{\beta},
\end{equation}
with $0<\gamma \leq 1$ given by Eq.~\eqref{eq:gamma2}.
In other words, the more molecules we fit into the cavity the more red-shifted the cavity becomes, due to self-consistently considering cavity-induced inter-molecular polarization effects.

How does this self-consistent redshift compare against the frequency shift in a Maxwellian standing-wave picture, introduced in Eq.~\eqref{eq:FabryPerotCavity}, where the influence of the matter is described via the (static) refractive index of the medium?
%That is, we have derived a change of the vacuum refractive index from first principles.
%We will discuss below how this self-consistent change of refractive index compares with the standard classical refractive index derived from macroscopic polarization.
%
For external driving fields, the optical wavelength $\Lambda$ of the Fabry-P\'erot resonator depends on the refractive index of the contained medium. 
In more detail, the resulting frequency shift can be related to the real part of the refractive index $n_\mathrm{r}$ via
\begin{equation}\label{eq:freq-shift-ref-index}
    \tilde{\omega}_{\beta}^{\rm M} = \frac{\omega_{\beta}}{n_{\rm r}} .
\end{equation}
The refractive index in a dilute (gaseous) medium depends on its number density $N/V$ and is given by~\cite{atkins2011molecular}
\begin{equation}\label{eq:ref-classical}
    n_r = \sqrt{\frac{4 \pi N \alpha_i}{V} +1}
\end{equation}
for bare matter.
Assuming that the mode volume $\mathcal{V}$ corresponds to the physical volume $V$ of the diffractive medium, we can use $\lambda^2$ from Eq.~\eqref{eq:lambda-def} and substitute into Eq.~\eqref{eq:freq-shift-ref-index} to write
\begin{equation}\label{eq:scaled-cav-explicit}
    \tilde{\omega}_{\beta}^{\rm M} = \omega_{\beta} \left(N \lambda^2 \alpha_i + 1 \right)^{-\frac{1}{2}},
\end{equation}
which depends on the bare single-molecule polarizability  $\alpha_i$ given in Eq.~\eqref{eq:polarizability}. 
Comparing the redshifts determined with Eqs.~\eqref{eq:selfconsistentfrequency} and \eqref{eq:scaled-cav-explicit}, respectively, we find 
\begin{eqnarray}
    \tilde{\omega}_{\beta}=\tilde{\omega}_{\beta}^{\rm M}. \label{eq:redshift_relation}
\end{eqnarray}
Notice that an identical expression for the red-shift was found in Ref. \citenum{fiechter_understanding_2024} using second order perturbation theory with respect to the coupling strength.
The dependency of the red-shift on bare matter quantities is a little surprising, since we have just seen in the Sec~\ref{sec:pol} that the cavity changes the molecular polarizabilities. Therefore, naively, one would expect that $\tilde{\alpha}_i$ and the corresponding $\tilde{n}_r$ should be considered in Eq.~\eqref{eq:scaled-cav-explicit} rather than bare matter properties. Our results show the opposite, i.e., at least within our effective harmonic electronic structure model, the detuning of the cavity is related to bare matter polarizabilities and thus cannot directly be used to access the cavity-induced polarizability changes. To what amount this harmonic result remains valid for more complex electronic structures will be investigated ab-initio in Sec.~\ref{sec:cbo-hf} for real three-dimensional CO$_2$ molecules. Overall, this "boring" result again highlights that the properties of a collectively-coupled ensemble are not as straightforward as one might think.

%In other words, the self-consistent red shift of $N$ molecules inside a cavity corresponds to an refractive index equivalent to one obtained from macroscopic considerations which disregard screening of neighboring molecules.

Notice that the above findings apply for dilute media. It would thus be interesting to extend our results to the condensed phase. For non-polar condensed systems one usually models the response of the molecular ensemble (refractive index) by the Clausius-Mossotti equation instead, which considers screening effects from the induced polarization. A priori it seems rather unlikely that the relation in Eq.~\eqref{eq:redshift_relation} still persists in the condensed phase. Any experimentally measured deviations from the Clausius-Mossotti relation could provide important microscopic insights to cavity-modified electronic changes and to distinguish the effective mode volume $\mathcal{V}$ from the physical volume of the cavity.~\cite{svendsen2023theory}
% \revDS{Do you agree with my statement?}
%\tbdel{Therefore, a priori, we would consider our harmonic model reasonable to describe cavity-modified changes of the polarizability in the gas phase, but not necessarily for the condensed phase, as we have already assumed when using the cavity-Hartree equations.}
%
%
This subtle difference might also be reflected in the experimentally observed resonance condition under collective strong coupling.~\cite{thomas2016ground,patrahau_direct_2024} That is, the ensemble might only respond coherently if the cavity is tuned on resonance with some infrared-active ro-vibrational transition of the molecules in the ensemble. The present model apriori assumes classical coherence between the molecules on the time-scale of the electrons.
%Finally, we note in our harmonic setup, both the red shift and the polarizability modification appear to be independent of the tuning of the cavity, i.e., there is no molecular resonance condition. However, this may be a consequence of apriori assuming $V=\mathcal{V}$, while the effective mode volume in fact could depend non-trivially on the cavity and the chemical ensemble. However, this aspect goes beyond the scope of this work.
%\tbdel{The effective single electron charge number $Z_e$ can be seen as a model parameter that can be fitted to match the experimentally observed cavity-induced de-tuning of $\tilde{\omega}_\beta=\gamma^2\omega_\beta$  for realistic molecules with known bare molecular polarizabilities $\alpha_i$.}
\subsection{Nuclear equations of motion}

In a next step, we determine the classical equations of motion for the nuclei evolving on the $l$-th collective electronic potential surface. Again by employing the Hellmann-Feynman theorem, we obtain the force acting on the $n$-th nucleus in molecule $i$ from Eq.~\eqref{eq:pf_dip_h} as
\begin{eqnarray}
M_n\ddot{R}_{in} &=& -\frac{\mathrm{d}}{\mathrm{d} R_{in}} V(\boldsymbol{R}_i) -k_e (R_{in}-\langle \hat{r}_i\rangle_l)\nonumber\\
&&+\omega_\beta\Bigg( q_\beta - \frac{X+\langle \hat{x}\rangle_l}{\omega_\beta}\Bigg)\lambda Z_n.  
\end{eqnarray}
Afterwards, we relate the forces to the transverse electric field from Eq.~\eqref{eq:eperpdef}, which yields the following local equation of motion for the $n$-th nucleus of molecule $i$:
\begin{align}
    M_n\ddot{R}_{in} \overset{\rm Eq. \eqref{eq:rietrans}}{=}
        &-\frac{d}{dR_{in}} V_i(\mathbf{R}_i)-k_e \bigg(R_{in}-\sum_{m=1}^{N_n}\frac{R_{im}}{N_n}\bigg) \nonumber\\
        &+\bigg(Z_n - \frac{Z_e}{N_n}\bigg)E_\perp. \label{eq:localdyn1}
\end{align}
Notice that the first line contains purely intra-molecular forces (that are present even without a cavity), whereas the second line corresponds to forces that are solely collective and induced by the cavity.
As a sanity check, we immediately find that for neutral atoms, where we have only a single nucleus and thus $N_n=1$ and $Z_e=Z_N$, the nuclear motion is not influenced by the cavity. The equations of motions given in Eqs. \eqref{eq:qddot} and \eqref{eq:localdyn1} will later be used in Sec. \ref{sec:CO2} for ab-initio MD simulations of an ensemble of harmonic CO$_2$ molecules. 
While for a purely harmonic system classical and quantum equations of motion agree, the two descriptions will start to deviate for anharmonic systems (due to anharmonicities in the electron-nucleus or nucleus-nucleus interaction) and one would have to consider, e.g., hierarchical equations of motion, at high level of accuracy \cite{tanimura1989time,akbari2012challenges}. 
Having access to the exact harmonic dynamic,  we can now assess different commonly employed approximations in the following section.
%Therefore we need to have a dedicated molecular model that is not just an adapated atomic model.

\begin{table*}[]
\caption{Comparison of position and field observables / equations of motions for different approximations. \footnote{
The self-consistent reference calculations are denoted (sc). Approximations which neglect the dipole-self energy or any modification of the electronic structure, are labeled (D) and (be), respectively.  
%Exact electronic expectation values and eom of nuclei and displacement coordinate from 1D harmonic CBOA for an arbitrary nuclei potential $V_i(\mathbf{R}_i)$ are compared against two common approximation, which neglect dipole-self energy (dse) terms and modification of the electronic structure (be), respectively.
The approximations modify Eq.~\eqref{eq:shifted-HO} as follows: $\nu_{1, i}^\mathrm{D} = -k_e \mathcal{R}_i + Z_e \lambda \omega_\beta q_\beta$ and $\nu_{1, i}^\mathrm{be} = -k_e \mathcal{R}_i$ with $\nu_2 = \sqrt{N_\mathrm{n} k_\mathrm{e}}$ for both cases.
We use $Z_n' = Z_n - \frac{Z_e}{N_n}$, the nuclear sums $\mathcal{R} = \sum_j^N \mathcal{R}_j = \sum_j^N \sum_m^{N_\mathrm{n}} R_{jm}$ and $X' = X - \frac{Z_e \lambda}{N_n} \mathcal{R} = \lambda \sum_j^N \sum_m^{N_\mathrm{n}} Z_m' R_{jm}$.
Furthermore, we have defined the nuclear potential such that we would get the nuclear equation of motion without cavity, i.e., $-\frac{\partial}{\partial R_{in}} W (\mathbf{R}) = - \frac{\partial}{\partial R_{in}} V (\mathbf{R}) - k_e \big( R_{in} - \frac{1}{N_\mathrm{n}} \mathcal{R}_i \big)$.
%In the displacemnt field case, the cavity frequency is red shifted by scaling with $\frac{2 \gamma^2 - 1}{\gamma^2} = 1 - \frac{N Z_e^2 \lambda^2}{N_\mathrm{n} k_e}$.
Note that for the case of a neutral atom ($Z_n = Z_e$ and $N_\mathrm{n} = 1$) the nuclear motion correctly decouples from the cavity for (sc) and (D) because $Z_n' = 0$ .
}}
\label{tab:comparison-approximation}
\begin{tblr}{lllll} %l
%\begin{tabular*}{\textwidth}{@{\extracolsep{\fill}}llllllll@{}}
\toprule
 Level &
   $\braket{\hat{r}_i}$ %\eqref{eq:r_i_base} 
   &
   $\braket{\hat{x}}$ %\eqref{eq:relationx}
   &
   $E_\perp$ %\eqref{eq:eperpdef}
   &
   Equation of motion \\
    %$\begin{aligned}&\textstyle M_n \ddot{R}_{in} \quad \eqref{eq:localdyn1} \\ 
%&\textstyle + \frac{\partial W (\mathbf{R})}{\partial R_{in}} \end{aligned}$
   %$\ddot{q}_\beta$ \eqref{eq:qddot-base} \\ 
\addlinespace 
\midrule
\textbf{sc} &
   $\begin{aligned}\textstyle \frac{1}{N_\mathrm{n}} \mathcal{R}_i- \frac{Z_e}{N_\mathrm{n} k_e} E_\perp^\mathrm{sc}\end{aligned}$ &
   $\begin{aligned}\textstyle &
  \gamma^2 \bigg[ \frac{1 - \gamma^{2}}{\gamma^2} \big( \omega_\beta q_\beta - X\big) -\frac{Z_e \lambda}{N_\mathrm{n}} \mathcal{R} \bigg]\end{aligned}$ &
   $\begin{aligned}\textstyle
   \gamma^2 \lambda (\omega_\beta q_\beta - X')\end{aligned}$ &
   $\begin{aligned}\textstyle M_n \ddot{R}_{in} &= \textstyle -\frac{\partial W (\mathbf{R})}{\partial R_{in}} + Z_n' E_\perp^\mathrm{sc}\\
   \ddot{q}_\beta &= - \gamma^2 \omega_\beta^2 q_{\beta} + \gamma^2 \omega_\beta X' \end{aligned}$ \\
   \addlinespace 
 \textbf{be} &
   $\begin{aligned}\textstyle \frac{1}{N_\mathrm{n}} \mathcal{R}_i\end{aligned}$ &
   $\begin{aligned}\textstyle - \frac{Z_e \lambda}{N_\mathrm{n}} \mathcal{R}\end{aligned}$ &
   $\begin{aligned}\textstyle \lambda \big( \omega_\beta q_\beta - X' \big)\end{aligned}$ &
   $\begin{aligned}\textstyle M_n \ddot{R}_{in} &= \textstyle -\frac{\partial W (\mathbf{R})}{\partial R_{in}} + Z_n E_\perp^\mathrm{be} \\
   \textstyle \ddot{q}_\beta &= -\omega_\beta^2 q_\beta + \omega_\beta X'\end{aligned}$ \\ 
\addlinespace 
 \textbf{D} &
   $\begin{aligned}\textstyle \frac{1}{N_\mathrm{n}} \mathcal{R}_i- \frac{Z_e}{N_\mathrm{n} k_e} E_\perp^\mathrm{D}\end{aligned}$ &
   $\begin{aligned}\textstyle \frac{1 - \gamma^{2}}{\gamma^2} \omega_\beta q_\beta - \frac{Z_e \lambda}{N_\mathrm{n}} \mathcal{R}\end{aligned}$ &
   $\begin{aligned}\textstyle \lambda \omega_\beta q_\beta\end{aligned}$ &
   $\begin{aligned} M_n \ddot{R}^\mathrm{D}_{in} &= \textstyle -\frac{\partial W (\mathbf{R})}{\partial R_{in}} + Z_n' E_\perp^\mathrm{D}\\
   \ddot{q}_\beta &= \textstyle - \frac{2 \gamma^{2} - 1}{\gamma^2} \omega_\beta^2 q_\beta + \omega_\beta X'\end{aligned}$ \\
\bottomrule
 \end{tblr}
\end{table*}

\begin{table*}[]
\caption{Comparison of the dressed single-molecule (local) polarizabilities $\tilde{\alpha}_i$ and the detuning of the cavity frequency $\tilde{\omega}_\beta$ for the self-consistent (sc) reference setup with respect to the displacement field-only (D) and the bare electron (be) approximation. 
%Notice, that the erroneous local polarizabilities for the (D) and (be) setup emerge for a perturbative as well as for a self-consistent inclusion of the external electric field perturbation to the respectively approximated electronic structure problems. 
%For the (sc) reference, the correct polarizability $\tilde{\alpha}_i$ is only found when including the external field self-consistently. 
}
\label{tab:comparison-pol-detuning}
\begin{tblr}{lll} %l
%\begin{tabular*}{\textwidth}{@{\extracolsep{\fill}}llllllll@{}}
\toprule
 Level &
   Local Polarizability  &
    Cavity Mode Frequency \\
    %$\begin{aligned}&\textstyle M_n \ddot{R}_{in} \quad \eqref{eq:localdyn1} \\ 
%&\textstyle + \frac{\partial W (\mathbf{R})}{\partial R_{in}} \end{aligned}$
   %$\ddot{q}_\beta$ \eqref{eq:qddot-base} \\ 
\addlinespace 
\midrule
 \textbf{sc} &
   $\tilde{\alpha}_i=\gamma^2(N, \lambda)\alpha_i %=\frac{1}{1+\frac{\lambda^2 Z_e^2 N}{N_n k_e}}  \frac{Z_{e}^{2}}{{N_\mathrm{n}} k_{e}}
   $ &
$\tilde{\omega}_\beta=\gamma(N, \lambda)\omega_\beta %=\sqrt{\frac{1}{1+\frac{\lambda^2 Z_e^2 N}{N_n k_e}}}  \omega_\beta  
$\\
\addlinespace 
 \textbf{be} &
    $\tilde{\alpha}_i^\mathrm{be}=\alpha_i %= \frac{Z_{e}^{2}}{{N_\mathrm{n}} k_{e}}
    $  &
   $\tilde{\omega}_\beta^\mathrm{be}= \omega_\beta
   $  \\ 
\addlinespace 
 \textbf{D} &
   $\tilde{\alpha}_i^\mathrm{D}=\alpha_i% = \frac{Z_{e}^{2}}{{N_\mathrm{n}} k_{e}}
   $ &
   $\tilde{\omega}_\beta^\mathrm{D}=\sqrt{2-1/\gamma^2(N,\lambda)}\omega_\beta %=\sqrt{1-\frac{\lambda^2 Z_e^2 N}{N_n k_e}}  \omega_\beta 
   $   \\
\bottomrule
 \end{tblr}
\end{table*}
% -----------
% Comparison different approximations
% -----------
\subsection{Common Approximations}\label{sec:approximations}

Next we want to compare our self-consistent results with two common approximations that are used to simplify the dressed electronic-structure problem for VSC. 
%problems (see \autoref{tab:comparison-approximation} and \autoref{tab:comparison-pol-detuning}):
%There are several levels of approximations from calculation the full self-consistent coupling of the electronic system to the cavity (see \autoref{tab:comparison-approximation}).
%One commonly employed is to include modification of electronic structure but no polarization, i.e., neglect the dipole self-energy terms and thus neglecting cavity mediated matter-matter interaction (both self-consistent electron-electron and electron-nuclei). \revJH{Give refs}
%Already investigated, due to the analytic nature of our harmonic model access to observables is straightforward.
The first approximation is to disregard any modification of the electronic structure and instead only consider the bare (uncoupled) electronic structure problem. We denote this approximation in the following as \textbf{bare electronic (be)} approximation.
In the literature this approximation is sometimes also called the ``$\mu^2$'' approximation where $\mu$ labels the sum over the electronic and nuclear dipole moments.\cite{fischer2022CavityalteredThermalIsomerization, bonini2022InitioLinearResponseApproach, fiechter_understanding_2024} 
Looking at the corresponding equation of motion for the nuclear and displacement field dynamics in \autoref{tab:comparison-approximation}, we immediately notice that non-physical effects emerge. For instance, the bare electronic approximation leads to a spurious coupling of neutral atoms to the cavity within the dipole approximation. That is, the atomic nuclei couple via their nuclear charge $Z_n$ to the field since the screening of the nuclear charge via the effective electron is missing.
Furthermore, the bare electronic approximation can neither account for the experimentally observed cavity mode renormalization (red-shift) nor for any (self-consistent) change of the molecular polarizabilities (see \autoref{tab:comparison-pol-detuning} for details). Notice that any further simplified model will also lack these fundamental features, e.g., the Tavis-Cummings model that disregards the electronic structure entirely under VSC.

%This assumption, obtained by dropping the quadratic dipole self-energy (dse) terms in Eq.~\eqref{eq:pf_dip_h} and only keeping $-\omega_{\beta} \hat{q}_{\beta} (\hat{X} + \hat{x})$, is intrinsic of various approximations such as the Tavis-Cummings model~\cite{sidler.ruggenthaler.ea_2022, ruggenthaler.sidler.ea_2023}. \revJH{Not so sure about the Tavis-Cummings thing. Hopfield is µ model but more extreme? Without dse we will call it linear}
%\tbdel{We would obtain the same result if we replaced the displacement field by an external electric field instead. In more detail, via $-\omega_{\beta} \hat{q}_{\beta} (\hat{X} + \hat{x}) \rightarrow -E_{\rm ext}(\hat{X}+\hat{x})|e|/\lambda$ we could just replace $\omega_{\beta} q_{\beta} \lambda \rightarrow E_{\rm ext} |e|$ in Eq.~\eqref{eq:localdipolenoselfenergy}. \revJH{why is this interesting?}
%We note that this is only possible because we have a harmonic potential, which becomes arbitrarily strong as we move away from the minimum of the binding potential. For more realistic (Coulombic) potentials, which approach a finite value for $|r_i| \rightarrow \infty$, such dipolar couplings would become problematic since no ground state would exist in an ab-initio description~\cite{rokaj_lightmatter_2018,schafer_relevance_2020}.}

A second common approximation is to discard  the quadratic dipole self-interaction terms in Eq.~\eqref{eq:pf_dip_h} for the electrons and nuclei. However, in contrast to the bare electronic approximation, one keeps the coupling to the displacement field $-\omega_{\beta} \hat{q}_{\beta} (\hat{X} + \hat{x})$. The resulting electronic structure problem cannot account for cavity-mediated polarizations according to Eq.~\eqref{eq:eperpdef}.
This simplification we denote as \textbf{displacement-field coupling (D)} approximation. It is commonly employed, e.g., for the coupling of the bare nuclear Tavis-Cummings model.~\cite{sidler.ruggenthaler.ea_2022, ruggenthaler.sidler.ea_2023} Disregarding the dipole self-energy term for the nuclei and electrons as shown in \autoref{tab:comparison-approximation} is at least consistent with the exact dynamics of Eq.~\eqref{eq:localdyn1} in the trivial case of an ensemble of neutral atoms.
%For lin the coefficients in our model Hamiltonian from \eqref{eq:shifted-HO} become $\nu_{1, i}^\mathrm{D}$ and $\nu_{2}$
%The details are provided in \autoref{tab:comparison-approximation}).
Furthermore, on a first glance the cavity-mode renormalization $\tilde{\omega}_\beta^{\rm D}$ (red shift) appears similar to the reference calculations (sc), at least for small collective couplings, i.e., for $\gamma^2\approx 1$. However, as we will see later, when looking at CO$_2$ molecules, the displacement-field coupling approximation suggests a qualitatively erroneous scaling behavior with respect to the collective coupling strength. Furthermore, 
we note that the displacement-field coupling approach is only possible because we have a harmonic electronic model, since the binding potential for the effective electron becomes arbitrarily strong as we move away from the minimum. For realistic Coulombic interactions, which approach a finite value for $|r_i| \rightarrow \infty$, such dipolar couplings are  non-physical since no ground state exists in the basis-set limit~\cite{rokaj_lightmatter_2018,schafer_relevance_2020,schafer_relevance_2020}. Moreover, neglecting the dipole self-energy term does not yield any changes of the molecular polarizabilities, in contrast to the exact self-consistent solution. 
%However, results from lin have to be interpreted with caution as for example modification of reaction barrier height vanished when including dse terms. \cite{fischer2022CavityalteredThermalIsomerization}
Thus, when neglecting the exact polarization of the electronic structure, non-physical effects might emerge and we lack any polarization mechanism that can affect the molecular ensemble under collective VSC.

\section{Example CO$_2$}\label{sec:CO2}
\subsection{Many CO$_2$ molecules}

As a concrete example we investigate an ensemble of $N$ one-dimensional CO$_2$ molecules. It is a simple, linear and charge-neutral molecule, which possesses an infra-red active and infra-red inactive vibrational mode. Furthermore, CO$_2$ under VSC has previously been studied ab-initio, using different levels of approximations~\cite{flick_cavity-correlated_2018, bonini2022InitioLinearResponseApproach, fischerQuantumChemistryApproach2024a}. However, so far, the dressed electronic structure of molecular ensembles has not be solved self-consistently. 
We first write the equation of motions for a system of one-dimensional CO$_2$ molecules inside a cavity with respect to its bare eigenmode coordinates $(\boldsymbol{\rho}_{\rm t},\boldsymbol{\rho}_{\rm s},\boldsymbol{\rho}_{\rm a})$ (given in \autoref{sec:appendix-uncoupled-co2}) using Eqs.~\eqref{eq:qddot} and \eqref{eq:localdyn1}.
For this purpose we determine the self-consistent transverse electric field in terms of $(\boldsymbol{\rho}_a, q_{\beta})$. We do so by using Eq.~\eqref{eq:relationx} in conjunction with Eq.~\eqref{eq:eperpdef}, such that 
\begin{align}\label{eq:selfconsistentelectricfield}
E_{\perp} = \lambda \gamma^2 \left(\omega_{\beta} q_{\beta} + \lambda \epsilon_{\rm a} \rho_{\rm a} \right),
\end{align}
where the collective asymmetric vibrational mode is defined as $\rho_{\rm a} = \sum_{i=1}^{N}\rho_{{\rm a},i}$ and
\begin{align}
\epsilon_{a} = \frac{\sqrt{2 M} {\left(Z_{C} - Z_{O}\right)}}{3 \, \sqrt{M_{C}} \sqrt{M_{O}}}.
\end{align}
We note that the results in this section are obtained for a system of neutral molecules, i.e., $2Z_O + Z_C = Z_e$.
For the charged molecules, we would have deviations such as a contribution from the translational eigenmode to the transverse electric field of the form $-\lambda^2 \gamma^2 \epsilon_{\rm t} \rho_{\rm t}$.
The equations of motion of the $i$-th CO$_2$ molecule under VSC in normal mode coordinates is then given by
\begin{align}
\ddot{\rho}_{{\rm t},i} &= 0,
\\
\ddot{\rho}_{{\rm s},i} &= - k_{\rm s} \rho_{{\rm s},i},\label{eq:singemoleculesymm}
\\
\ddot{\rho}_{{\rm a},i} &= - k_{\rm a} \rho_{{\rm a},i} - \left(\epsilon_{\rm a} \lambda \rho_{\rm a} + \omega_{\beta} q_{\beta} \right) \epsilon_{\rm a} \gamma^2 \lambda, \label{eq:singemoleculeasymm}
\\
\ddot{q}_{\beta} &= - \gamma^2 \omega_{\beta} \epsilon_{\rm a} \lambda \rho_{\rm a} -\underbrace{\omega_{\beta}^2 \gamma^2}_{= \tilde{\omega}^2_{\beta}} q_{\beta}. \label{eq:singlemoleculephoton}
\end{align}
These equations provide valuable insights even without propagating them. We first note that only the asymmetric mode $\rho_{{\rm a},i}$ couples to the cavity. This may not seem surprising at first sight, since it is well established that symmetric stretch modes are not infra-red active in linear spectroscopy, which can be shown with perturbative theoretical arguments.~\cite{atkins_molecular_2011}
However, as we have seen for the polarizabilities under collective strong coupling, a perturbative picture may be misleading and thus a priori it is not clear why the symmetric and translational modes should not couple to the cavity as well for our self-consistently solved harmonic model. The sole coupling of the asymmetric mode becomes only evident from Eq.~\eqref{eq:singemoleculesymm}. For real (anharmonic) molecules with complex electronic structures, small corrections seem possible. That is, also other modes could self-consistently couple to the cavity. This is clearly an interesting question for future investigations.

In the following discussion, we can safely ignore the uncoupled translational and symmetric modes. If we look at the single-molecule Eq.~\eqref{eq:singemoleculeasymm}, the collective motion acts similar to an external force. Thus we find a feedback effect from the collective on the single-molecule vibrational motion. 
%\tbdel{This becomes specifically relevant in the case of a thermal ensemble, where the higher the frequency of the mode, the less it gets affected for a fixed temperature. Thus we find that the softening of the cavity and at the same time the stiffening of the collective asymmetric modes can lead to a change in the external force on the individual molecule. That is, the effective temperature on the individual molecules can be changed. How much, will then depend on the molecular structure, the number of molecules and the properties of the cavity, i.e., the coupling constant $\lambda$. We note, however, that this fully harmonic picture might be too simplistic, specifically in the large-$N$ limit. We discuss this in Sec.~\ref{sec:largeNlimit} below.}
%
The corresponding equation of motion of the collective asymmetric normal mode is then
\begin{align}
\ddot{\rho}_{{\rm a}} &= - \underbrace{\left(k_{\rm a} %-
+ N \epsilon_{\rm a}^2 \lambda^2 \gamma^2 \right)}_{= \tilde{k}_a} \rho_{{\rm a}} - \frac{N \epsilon_{\rm a}  \lambda }{\omega_{\beta}} \tilde{\omega}^2_{\beta} q_{\beta} , \label{eq:collectiveasymmmode}
%\\
%\ddot{q}_{\beta} &= - \tilde{\omega}^2_{\beta} q_{\beta} - \gamma^2 \omega_{\beta} \epsilon_{\rm a} \lambda \rho_{\rm a}. \label{eq:collectivephotonmode}
\end{align}
Here we see a change in the collective force constant $k_a \rightarrow \tilde{k}_{a} = \left(k_a + N \epsilon_{\rm a}^2 \lambda^2 \gamma^2 \right)$ of the collective infrared mode $\rho_{\rm a}$.
That is, besides the red shift of the cavity mode we find an accompanying stiffening of the collective vibrational mode.

\begin{figure}
    \centering
    \includegraphics{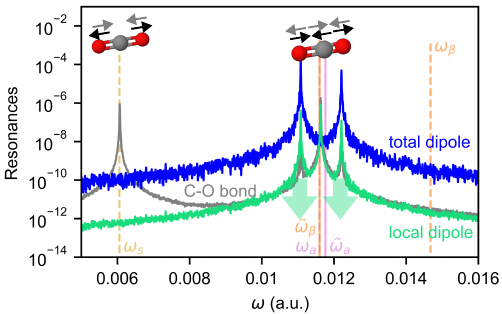}
    \caption{Molecular-dynamics simulation for one-dimensional and harmonic CO$_2$ molecules with the cavity frequency tuned to the asymmetric vibration yields resonance spectrum with Rabi split in agreement to normal mode representation from \autoref{sec:CO2}. 
    Solid plots are C-O bond (grey), total dipole (blue) and local dipole (green) resonances calculated from molecular-dynamics autocorrelation functions; for details see appendix~\ref{sec:appendix-numerical_details}. Vertical lines mark system dependent frequencies: symmetric vibration $\sqrt{k_s}$, and asymmetric vibration $\sqrt{k_a}$ (purple) and cavity $\omega_\beta$ (orange). While the bare system frequencies are dashed, renormalized frequencies due to coupling like $\tilde{k}_a$ and $\tilde{\omega}_\beta$ are drawn solid. Note that here self-consistent coupling to a cavity mode follows traditional infra-red selection rules, i.e., the symmetric vibration does not couple to the cavity. 
    Furthermore, for small number of molecules and high simulation strength, local polaritons can be seen. However, when increasing the number of molecules while keeping the Rabi splitting (or $\lambda_\mathrm{col}$) constant, their intensities vanish (indicated by green arrows) leaving only the peak for the asymmetric vibration. }
    \label{fig:numerical-sim}
\end{figure}

Finally, we simulate the nuclear dynamics of an ensemble of CO$_2$ molecules having the cavity tuned on resonance with the infra-red active vibrational mode (see \autoref{fig:numerical-sim}). The simulations are performed in Cartesian coordinates, which is an excellent consistency check for the normal mode discussion as well to verify the analytic results of the cavity renormalization. The simulations rely on the analytic forces given by Eq.~\eqref{eq:localdyn1} together with a thermostat (described in detail in \autoref{sec:appendix-numerical_details}). This allows for a highly efficient implementation, since there is no need anymore to solve the electronic-structure problem numerically.
The vibrational spectra can be determined locally (green) and collectively (blue) by evaluating the corresponding dipole fluctuations using standard techniques from molecular dynamics (see e.g. Ref.~\citenum{sidler_unraveling_2024}).
Besides confirming the analytic red-shift of the cavity frequency $\omega_\beta$, we observe a Rabi split in both collective as well as local dipole moments. 
However, the local polaritons in the harmonic CO$_2$ system vanish when increasing the number of molecules, while keeping the collective Rabi-splitting fixed (green arrows).
This we also anticipate analytically in the local normal mode picture of Eq.~\eqref{eq:singemoleculeasymm} when applying the Tavis-Cummings scaling argument given in Eq.~\eqref{eq:collective-coupling-strength}. Thus, the vibrational modes seem locally unaffected by the cavity in the harmonic model, despite having modified polarizabilities. It seems that a more complex electronic/nuclear structure is required (i.e., anharmonic) to obtain such (small, but finite) effects on the nuclear dynamics. For example, see the numerical discussion of an ensemble of Shin-Metiu molecules  in Ref.~\citenum{sidler_unraveling_2024} that allows for a non-vanishing polarization pattern under collective VSC (polarization glass). 

\subsection{Comparison to ab initio results\label{sec:cbo-hf}}
%\revDS{results and interpration incomplete...}
For the harmonic model we could surprisingly relate the self-consistent red shift of the cavity frequency to its bare matter refractive index, despite observing a cavity-induced change of the molecular polarizabilities.
To verify if Eq.~\eqref{eq:redshift_relation} still holds for complex electronic structures, we calculate the self-consistent vibro-polaritonic infra-red spectra of one CO$_2$ molecule coupled to a single cavity mode within the cavity-Born-Oppenheimer Hartree-Fock (cBO-HF) approximation~\cite{schnappinger2023ab}.
Note that perturbative approaches to determine the cavity-mode renormalization\cite{fischerQuantumChemistryApproach2024a, fiechter_understanding_2024} recover the same red shift as our harmonic model in Eq.~\eqref{eq:scaled-cav-explicit}. Thus the results of the following analyis are directly transferable.

\begin{figure*}
    \centering
    \includegraphics{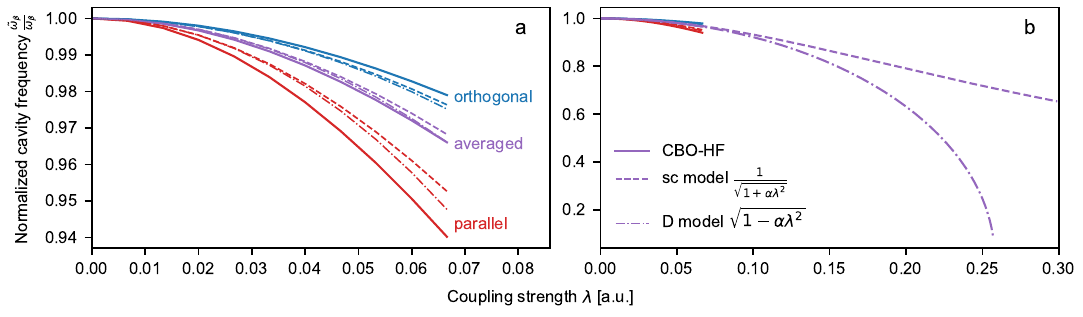}
    \caption{a) Red shift (bold lines) of the cavity frequency from ab-initio cBO-HF calculation for a single realistic CO$_2$ molecule coupled to a cavity mode aligned parallel (red) or orthogonal (blue) to the bond axis is compared to analytical expressions from our model. The self-consistent (sc) model (dashed lines) and the displacement-field coupled (D) model (dot-dashed lines) red shifts use the bare matter polarizability and are calculated based on \autoref{tab:comparison-pol-detuning}. 
    %Note that calculating a red shift via cBO-HF relies on the existence of a mode with a clear photonic character for the whole coupling strength regime. Thus detuning from IR active matter modes, i.e., orthogonal to asymmetric mode or tuned to symmetric mode is needed for practical purposes. %The red shift is stronger when the molecule is aligned parallel (z-axis) to the cavity mode. 
    The analytical expressions based on the corresponding bare matter polarizability components $\alpha_{xx}$ and $\alpha_{zz}$ both over- and under-estimate the red shift for the orthogonal and parallel setup, respectively. After spatial averaging good agreement is found thanks to error cancellation. The coupling strength regime shown here corresponds to Rabi splittings up to a few hundred cm$^{-1}$ and covers the chemically relevant regime (based on typical polarizability densities). b) Increasing the coupling further reveals the
    non-physical behavior of the displacement-field coupling (\textbf{D}) approximation in the ultra-strong coupling regime.
    }
    \label{fig:sup-cav-frequency-scaled}
\end{figure*}

In \autoref{fig:sup-cav-frequency-scaled}, different cavity-induced redshifts are compared for a single CO$_2$ molecule with respect to different couplings $\lambda$ by means of normalized cavity frequencies $\tilde{\omega}_\beta / \omega_\beta$. The reference data is shown in solid lines, whereas dashed lines correspond to the self-consistent harmonic electronic model description $\tilde{\omega}_\beta^{\rm sc}$. The qualitatively different physics of the displacement-field coupling model $\tilde{\omega}_\beta^{\rm lin}$ is shown by dot-dashed lines. %and compare against the scaled cavity frequency from the refractive index (equivalent to dividing \eqref{eq:scaled-cav-explicit} by $\omega_\beta$ and comparing right and left side).
Note that for the reference cBO-HF results of $\tilde{\omega}_\beta (\lambda)$, the cavity mode needed to be detuned from any IR-active vibration to avoid Rabi splitting and preserve its photonic character in the eigenvector.

Comparing the redshifts for different molecular orientations with respect to the cavity, we find that the self-consistent harmonic model either over or underestimates the red shift of the cavity with respect to the reference. The deviations of the harmonic self-consistent model become larger, the more polarizable the real molecule is (i.e., parallel alignment).
Or alternatively, for weakly polarizable molecules and low coupling strength the analytic red shift in Eq. \eqref{eq:scaled-cav-explicit} reproduces the ab-intio results very well.
This also explains the almost perfect agreement in Ref.~\citenum{fiechter_understanding_2024} of the coupled spectra obtained from perturbation theory for a HF molecule, because its polarizability is small compared to CO$_2$.
% Therefore, those approaches suffer from the same deficiencies when it comes to larger couplings strength \citenum{fiechter_understanding_2024}
% This also indicates that a perturbative treatment of the cavity-mode renormalization\cite{fiechter_understanding_2024} is expected to become unreliable, the more polarizable the molecules are.  Therefore,  almost perfect agreement between Eq. \eqref{eq:scaled-cav-explicit} and ab-initio cBO-HF simulations is found for hydrogen fluoride.\cite{fiechter_understanding_2024}  %for orthogonal and parallel orientation (w.r.t. to the molecular axis which also is the largest component of the polarizability tensor).
Interestingly, the spatial averaging for an ensemble of randomly oriented molecules leads to significant error-cancellation and thus the analytic model and numerics agree relatively well. 
In more detail, the spatial averaging is applied with 2/3 of the molecules being orthogonal and 1/3 parallel.
The observed error cancellation suggests that our analytic redshift model might be applicable with relatively high accuracy to experimental data. Moreover, it could also be the reason why the standard Maxwellian approach $\tilde{\omega}_\beta^{\rm M}$ seems sufficient to predict cavity-induced detunings, even-though its small deviations could provide important physical insights into cavity-induced microscopic changes. 
%Finally, we note that the reference calculations were performed for a single molecule and then averaged a posteriori to mimic the random oriented molecular ensemble. However, this separation might be too simplistic and one would rather need to solve the cH equations for an ensemble of randomly oriented molecules instead to have an accurate reference picture. However, this goes beyond the current capabilities of the implementation of Ref. \citenum{schnappinger2023ab} and we leave it for future work.
%The analytical expression for the normalization of the cavity frequency in the case of the linear approximation (i.e., neglecting dse terms) is taken from \autoref{tab:comparison-approximation} as
%\begin{equation}\label{eq:lin-red-shift}
%\frac{\tilde{\omega}_\beta}{\omega_\beta} = \sqrt{1 - \alpha^\mathrm{lin}_i \lambda^2} 
%\end{equation}
%where we have used the fact that $\alpha^\mathrm{lin}_i$ is the same as the bare matter polarizability $\alpha_i$ from \eqref{eq:polarizability}.

%\revDS{I do not understand the next part. I'm confused. Is it not all trivial and can safely be removed?}
%Polarizability scales with $\gamma^2$ from analytical model, component along polarizability.
%From above we have identified
%$$
%\gamma = \frac{1}{3 \omega_\beta} \left(2 \tilde{\omega}_\beta^\perp + \tilde{\omega}_\beta^\parallel \right)
%$$
%and use this to calculate the polarizability as $\tilde{\alpha}_i = \gamma^2 \alpha_i$ where we can calculate the z-component as $\tilde{\alpha}_i^z = \gamma_\parallel^2 \alpha_i^z$ with $\gamma_\parallel =  \frac{\tilde{\omega}_\beta^\parallel}{\omega_\beta}$.
%However, 'semi-analytical' expression for average polarizability is tricky as this value scales different for orthogonal polarization of cavity while the calculated $\gamma$ value is the same.
%($\gamma$ is calculated using both frequency shifts for orthogonal and parallel setups.)
%Thus we cannot conclude that the difference in the bottom plot is only due to anharmonicities.
%\revJH{Derivations from CBO-HF cancel which makes interpretation dfficult. Effect of orientation future research}

\section{Conclusion}
Let us collect the major results that we have obtained. Thanks to the harmonic design of our model, the dressed electronic structure problem of an ensemble of molecules in an optical cavity can be solved analytically in the dilute gas limit. Our setup does not impose any approximations on the dipole self-energy term, thus allowing for intermolecular polarization effects. This provides  unique non-perturbative insights in the resulting scaling behavior of large molecular ensembles. 
Most notable, the self-consistent inclusion of an externally applied (static) electric field reveals a cavity-modified molecular polarizability, which is local in nature, but depends on the collective coupling strength. Therefore, this local polarization mechanism can persists in the thermodynamic limit. In contrast, a perturbative calculation of the polarizability would lead to a qualitatively different scaling behavior with vanishing local effects in the thermodynamic limit. Therefore, we conclude that it is vital to accurately (self-consistently) account for the delicate nature and changes of the cavity-induced local dipole-dipole interaction to recover the correct physics. Perturbation theory seems insufficient for this purpose and can even introduce qualitatively wrong scaling behavior. However, surprisingly, the exact (self-consistent)  polarizability was recovered perturbatively for our harmonic ensemble, when assuming ad-hoc single-molecular strong coupling conditions instead! This fundamental theoretical observation has far reaching implications to justify the broad applicability of the numerous recently developed ab-initio simulation methods, which are naturally restricted to single or few-molecular strong coupling situations due to their computational load. 

When investigating  the dynamics of the nuclei and displacement field coordinates for our analytical model, we also find an analytic expression for the renormalization of the cavity mode frequency, showing its relation to the bare matter refractive index of a dilute gas. The red-shift in the cavity thus appears to not be affected by the aforementioned changes of the local polarizability. Furthermore, when investigating the dynamics of the nuclei with a normal-mode analysis we find that despite the self-consistent treatment of the electronic structure problem, only the infra-red active  modes couple to the cavity. Comparing our results with different common approximations reveals not only for the polarizability and red-shift a qualitatively different physics, but also for the local equations of motions. Eventually, we compare the analytic red-shift formula of our harmonic model with the self-consistently determined value of one realistic CO$_2$ molecule, with particularly very good agreement for randomly oriented molecular ensembles. This indicates that our harmonic model is indeed capable of capturing relevant properties of real chemical system.

However, above results are clearly not the end of the story. For example, the locally modified polarizability is expected to alter the local fluctuations of the electronic structure. Consequently, the dispersion interactions and thus the van der Waals forces between the molecules should be modified by the cavity as already suggested by experimental evidence.~\cite{patrahau_direct_2024} Moreover, it will also be interesting to investigate how the subtle  interplay of local vs. collective properties behaves in the condensed phase. As a straightforward consequence, we expect the standard Clausius-Mossotti relations to be modified under collective VSC, i.e., the relation of the macroscopic dielectric constant to the single molecular polarizabilities might be altered in a non-trivial fashion due to self-consistent feedback effects. In addition, we highlight two other important properties of VSC that were not accessible with  our idealized harmonic ensemble model:  There is theoretical evidence for a polarization glass-like  frustration effects~\cite{sidler_unraveling_2024, sidler_connection_2024} as well as for a strong (experimentally detectable) resonance condition in cavity-modified reaction rates.~\cite{thomas2016ground,patrahau_direct_2024} However, we consider our model a first analytic step towards a better justification of recently developed rate theories\cite{wang_roadmap_2021,yang_quantum_2021,lindoy_resonant_2022,cao_generalized_2022,ying_resonance_2023,sun2022suppression,sun2023modification,ying_resonance_2024,anderson_mechanism_2023} for optical cavities that often rely on the applicability of the single-molecular strong coupling hypothesis under collective VSC. At least for the molecular polarizability,  we could now confirm this hypothesis ab-initio.   
We therefore expect that our self-consistent model will provide further valuable insights under various conditions (e.g. thermal equilibrium), that will help to unravel the mysteries of VSC.

\begin{acknowledgments}
\vspace{10pt}
JH and DS contributed equally to this work.
This work was made possible through the support of the RouTe Project (13N14839), financed by the Federal Ministry of Education and Research (Bundesministerium für Bildung und Forschung (BMBF)) and supported by the European Research Council (ERC-2015-AdG694097), the Cluster of Excellence “CUI: Advanced Imaging of Matter” of the Deutsche Forschungsgemeinschaft (DFG), EXC 2056, project ID 390715994 and the Grupos Consolidados (IT1453-22). 
The Flatiron Institute is a division of the Simons Foundation.
\end{acknowledgments}

%\begin{authorcontributions}
% --------------------------------------------
% APPENDIX
% --------------------------------------------
\appendix

\section{Derivation of Single Molecular Polarizabilities \label{sec:polder}}
Here we briefly give details to the derivation of the two quintessential relations for the polarizabilities shown in Eqs. \eqref{eq:coll_polarizability_sc_lim_loc} and \eqref{eq:wrong_scaling}. The self-consistently, determined local polarization for a collective external field-perturbation is defined as follows,

\begin{eqnarray}
   \tilde{\alpha}_i = \frac{\partial \braket{d_i}}{\partial E_\mathrm{ext}} \ \mathrm{with } \ \hat{H}_{\rm tot}=\hat{H} - \hat{d} E_\mathrm{ext}
\end{eqnarray}
Including $\hat{d_i} E_\mathrm{ext}=Z_e\hat{r_i} E_\mathrm{ext}$ explicitly when solving Eq. \eqref{eq:local_shift_osc}, we find for the local polarizability,

\begin{eqnarray}
   \tilde{\alpha}_i = Z_e\frac{\partial \braket{\hat{r}_i}}{\partial E_\mathrm{ext}} %= -\frac{Z_e}{\nu_2^2} \frac{\partial \nu_{1,i}(E_\mathrm{ext}) }{\partial E_\mathrm{ext}}
\end{eqnarray}
with
\begin{widetext}
\begin{eqnarray}
    \braket{\hat{r}_i}(E_\mathrm{ext})&=&-\frac{\nu_{1,i}(E_\mathrm{ext})}{\nu_2^2} \overset{\rm Eq.\ \eqref{eq:r_i}}{=}\frac{1}{N_n}\bigg(\sum_{n=1}^{N_n}R_{in} +\frac{\lambda Z_e }{ k_e}\Big(X +\langle x\rangle_0(E_\mathrm{ext})-  \omega_\beta q_\beta\Big)+\frac{Z_e}{k_e}E_\mathrm{ext}\bigg)\\
    \braket{x}_0(E_\mathrm{ext}) &\overset{\rm Eq.\ \eqref{eq:relationx}}{=} &\big(1 - \gamma^{2}(N, \lambda) \big) \bigg[
        - \frac{k_e}{\lambda Z_e N} \sum_{i}^N \sum_{n=1}^{N_\mathrm{n}} R_{i n}
        - X 
        + \omega_\beta q_\beta - \frac{1}{\lambda} E_\mathrm{ext}
    \bigg]
\end{eqnarray}
\end{widetext}
which leads to
\begin{eqnarray}
   \tilde{\alpha}_i& = &Z_e\frac{\partial \braket{\hat{r}_i}}{\partial E_\mathrm{ext}} \\
   &=&\frac{Z_e^2}{N_n k_e}(1 -(1-\gamma^2(N, \lambda)))=\alpha_i \gamma^2(N, \lambda)
\end{eqnarray}
and thus proves the cavity induced local polarizability change in Eq. \eqref{eq:coll_polarizability_sc_lim_loc}, when considering self-consistently the impact of the external electric field. The other self-consistent (local and non-local) polarizabilities introduced in the manuscript follow from equivalent derivations.

When calculating the local polarizability perturbatively, as it is common practice in quantum chemistry, we find
for
\begin{eqnarray}
     \tilde{\alpha}^{\rm pert}_i= -2\sum_{l\neq 0} \frac{\bra{0}Z_e \hat{r}_i\ket{l}\bra{l}Z_e \hat{r}_i\ket{0}}{\epsilon_i^0-\epsilon_i^l}
\end{eqnarray}
using that 
\begin{eqnarray}
    \bra{0} \hat{r}_i\ket{l} &=& \bra{0}\hat{U}_i^\dagger \hat{U}_i\hat{r}_i\hat{U}_i^\dagger \hat{U}_i \ket{l}=\bra{0^\prime} \hat{r}_i+\braket{r}_0\ket{l^\prime}\\
    &=&\bra{0^\prime} \hat{r}_i\ket{l^\prime}
\end{eqnarray}
where the $^\prime$ indicates a bare (no shift) quantum harmonic oscillator wave-function $\ket{\psi^\prime}=\hat{U}_i \ket{\psi}$. Consequently, only the first excited state has a non-zero contribution to the transition dipole element and we recover the dressed perturbative solution for the local polarizability as,\cite{atkins2011molecular}
\begin{eqnarray}
     \tilde{\alpha}^{\rm pert}_i&=& \frac{2 Z_e^2}{\nu_2}  \bra{0^\prime}\hat{r}_i\ket{1^\prime} \bra{1^\prime} \hat{r}_i\ket{0^\prime}\\
     &=&\frac{2 Z_e^2}{\nu_2}\bigg(\frac{1}{\sqrt{2 \nu_2}}\bigg)^2\\
     &=&\frac{Z_e^2}{\lambda^2 Z_e^2+N_n k_e}=\alpha_i \gamma^2(1, \lambda),
\end{eqnarray}
which leads to the erroneous perturbative scaling relation summarized in Eq. \eqref{eq:wrong_scaling}. The generalization to the collective case is trivial, yielding $\tilde{\alpha}^{\rm pert}=N\tilde{\alpha}^{\rm pert}_i$.

\section{Uncoupled CO$_2$ Molecule}\label{sec:appendix-uncoupled-co2}
Next, we want to write the equation of motion of the nuclei in such a way that the influence, i.e., the coupling, of the cavity can be easily observed.
The vibrations of a CO$_2$ molecule can be approximated by the classical motion of the nuclei each interacting harmonically with its neighbor via the potential
\begin{equation}\label{eq:harmonicnuclei}
W_i(\mathbf{R}_i) =  \frac{k_n}{2}(R_{i1}-R_{i2})^2 +\frac{k_n}{2}(R_{i2}-R_{i3})^2 \\
\end{equation}

For the bare molecules we can perform a simple normal mode expansion.
From the classical equations of motion for the bare molecules
\begin{eqnarray}
M_O \ddot{R}_{i1} &=& -k_n (R_{i1}-R_{i2}),\\
M_C \ddot{R}_{i2} &=& k_n (R_{i1}-R_{i2})-k_n (R_{i2}-R_{i3}),\\
M_O \ddot{R}_{i3} &=& k_n (R_{i2}-R_{i3}),
\end{eqnarray}
the following three normal-mode coordinates, denoted respectively as translational, symmetric and asymmetric, suggest themselves~\cite{atkins2011molecular}
\begin{eqnarray}
\rho_{{\rm t},i}&=&\frac{1}{\sqrt{M}}\left(M_O R_{i1}+M_C R_{i2}+M_O R_{i3}\right),\\
\rho_{{\rm s},i}&=&\frac{\sqrt{M_O}}{\sqrt{2}}\left( R_{i1}-R_{i3}\right),\\
\rho_{{\rm a},i}&=&\frac{\sqrt{M_O M_C}}{\sqrt{2 M}} \left(R_{i1}-2 R_{i2}+R_{i3}\right),
\end{eqnarray}
with $M=2M_O+M_C$. 
Their corresponding equation of motions are given by
\begin{align}
\ddot{\rho}_{{\rm t},i} &= 0,\label{eq:baretrans} \\
\ddot{\rho}_{{\rm s},i} &= - \underbrace{\frac{k_n}{M_O}}_{= k_{\rm s}} \rho_{{\rm s},i}, \\
\ddot{\rho}_{{\rm a},i} &= - \underbrace{\frac{M k_n}{M_O M_C}}_{= k_{\rm a}} \rho_{{\rm a},i}. \label{eq:bareasym}
\end{align}
As the translational mode describes the motion of the center of mass of the molecule it has 0 eigenfrequency.

To have the same equations of motions for our single-electron molecule outside the cavity we are free to make the choice for the nuclear potential as
\begin{equation}\label{eq:nuclei-potential-total-co2}
    V_i (\mathbf{R}_i) = W_i (\mathbf{R}_i) - \frac{k_e}{2} \Bigg(\sum_{n=1}^{3}  R_{in}^2 -  \frac{1}{3}\bigg(\sum_{n=1}^{3}R_{in}\bigg)^2 \Bigg) .
\end{equation}
That is, for the out of cavity case ($\lambda = 0$), which we also call ``bare matter'', \eqref{eq:localdyn1} becomes
\begin{align}
    M_n \ddot{R}_{in}^\mathrm{bare} &= -\frac{d}{dR_{in}} V_i(\mathbf{R}_i)-k_e \bigg(R_{in}-\sum_{m=1}^{N_n}\frac{R_{im}}{N_n}\bigg) \notag \\
    &=- \frac{\partial}{\partial R_{in}} W_i (\mathbf{R}_i)
\end{align}
and
\begin{equation}
    \braket{r}_i^\mathrm{bare} = \frac{1}{N_\mathrm{n}} \sum_i^N R_{in}
\end{equation}

\section{Numerical Details}\label{sec:appendix-numerical_details}
Molecular dynamics and spectra calculation follows the process from \cite{sidler_unraveling_2024} to which we refer the reader for more details.
The code was adapted to our model of harmonic one-dimensional CO$_2$ molecules where analytical force calculation reduces the computational cost.
We employ classical canonical equilibrium conditions by weakly coupling our ensemble 
to a thermal bath at temperature $T$.
The resulting classical Langevin equations of motion are given by,
\begin{align}
M_n\ddot{R}_{in} &= 
    -\frac{\partial W (\mathbf{R})}{\partial R_{in}} + Z_n' E_\perp^\mathrm{sc} \notag \\
    &\quad - \gamma_T M_n\dot{R}_{in} + \sqrt{2 M_n \gamma_T k_B T}S_{in} \\
\ddot{q}_{\beta} &= 
    -\omega_\beta^2 q_\beta +\omega_\beta (X+\langle x\rangle) \notag \\
    &\quad - \gamma_T \dot{q}_{\alpha} + \sqrt{2 \gamma_T k_B T}S_{\beta}\\
    \langle S_{in}(t)\rangle &=0 , \quad \langle S_{in}(t)S_{in}(t^\prime)\rangle = \delta(t-t^\prime)
\end{align}
with friction coefficient $\gamma_T$.
Atomic units were used throughout the calculations such that $M_C = 21874$ a.u. and $M_O = 29166$ a.u.
We chose the unit charge $Z_C = 1/2 Z_O = 1$ and $Z_e = Z_C + 2 Z_O$ and $k_e = 1$ while $k_n$ was calculated from cBO-HF asymmetric stretch ($\sqrt{k_a} = 2538$~cm$^{-1}$~$= 0.0116$ H) via the bare matter relation from \eqref{eq:bareasym}.
The classical Langevin equations of motion were propagated numerically using the scheme of Ref. \citenum{bussi2007AccurateSamplingUsing} with a time step $\delta t = 20$.
Trajectories were simulated over 200000 time-steps for $N=20$ molecules for spectra calculations.
The nuclei were initialized randomly distributed in the vicinity of the ground-state.
Thermostating parameters were set to $k_B T = 10^{-3}$ with low friction coefficient $\gamma_T = 0.5 \times 10^{-5}$ (underdamped regime).
%Notice that the parametrization of the molecule and the temperature was chosen such that non-adiabatic coupling effects should not play a role, i.e., the groundstate cBOA approximation is valid. Furthermore, the temperature was chosen small enough that the thermal broadening does not interfere with our spectral interpretation. The vibrational absorption spectra was calculated using the power spectra method in Ref. 14 with a Blackman filter window15 averaged over 33 overlapping trajectory windows containing 4096 time steps, each of them shifted by 1/3 of the window size. While for the global absorption spectrum the total dipole (electronic + nuclear contribution) were post-processed accordingly, we did the same for each individual molecular dipole in case of the local spectra calculation instead.
%Afterwards, the summation over all local spectra was taken yielding bold lines in the spectral figures.

Vibrational cBO Hartree Fock (cBO-HF) calculations were done as described in \cite{schnappinger2023ab} for a single CO$_2$ molecule with its bond axis oriented parallel and orthogonal to a single cavity mode tuned on resonance to the symmetric matter vibration (1496$^{-1}$~cm).
The cBO-HF ansatz have been implemented in the Psi4NumPy environment~\cite{Smith2018-tu}, which is an extension of the PSI4~\cite{Smith2020-kq} electronic structure package. 
All calculations were performed using the aug-cc-pVDZ basis set~\cite{Kendall1992-wu} and all geometries were pre-optimized at the Hartree-Fock level of theory. 
It should be noted that only the situation where the bond axis of CO$_2$ is oriented orthogonal to a single cavity mode is a real minimum, and the parallel situation represents a transition state along the rotation with respect to the polarization axis~\cite{Schnappinger2024-vt}.
%The CBO-HF polarizabilities were calculated using an adapted version of the Coupled Perturbed Hartree-Fock formalism~\cite{Helgaker1988-if} and will be the subject of a subsequent publication. 
The vibrational cBO-HF calculations were performed in a reproducible environment using the Nix package manager together with NixOS-QChem \cite{nix} (commit f5dad404) and Nixpkgs (nixpkgs, 22.11, commit 594ef126).

% --------------------------------------------------------------------
% --------------------------------------------------------------------
% --------------------------------------------------------------------

\bibliography{manuscript}

\end{document}